\documentclass[prd,aps,floats,draft,showpacs]{revtex4}

\def\lsim{\mathrel{\mathop
  {\hbox{\lower0.5ex\hbox{$\sim$}\kern-0.8em\lower-0.7ex\hbox{$<$}}}}}
\def\gsim{\mathrel{\mathop
  {\hbox{\lower0.5ex\hbox{$\sim$}\kern-0.8em\lower-0.7ex\hbox{$>$}}}}}

\begin{document}

\input epsf

\newcommand{\EMI}{\approx} % equal modulo a multiple of the identity
\newcommand{\HIGL}{\Phi_L} % Lattice Higgs field in matrix form
\newcommand{\higl}{\phi_{L}} % Lattice Higgs field in component form 
\newcommand{\INFL}{\chi_L} % Lattice Inflaton 
\newcommand{\Higg}{\Phi} % Continuum Higs field in matrix form
\newcommand{\higg}{\phi} % Continuum Higs field in component form 
\newcommand{\infl}{\chi} % Continuum inflaton form
\newcommand{\HINFK}{nose} % Continuum k momentum Higgs parallel-inflaton vector  
\newcommand{\II}{{\mathbf I}}
\newcommand{\FQ}{{\cal F}}
\newcommand{\Lam}{\Lambda} % The Lattice
\newcommand{\DCOV}{{\cal D}} % Lattice covariant derivative
\newcommand{\DBCOV}{\overline{\cal D}}  % Adjoint  Lattice covariant derivative
\newcommand{\trh}{T_{\rm rh}}
\newcommand{\teff}{T_{\rm eff}}
\newcommand{\mueff}{\mu_{\rm eff}}
\newcommand{\delcp}{\delta_{_{\rm CP}}}
\newcommand{\mh}{m_{_{\rm H}}}
\newcommand{\mw}{m_{_{\rm W}}}
\newcommand{\gw}{g_{_{\rm W}}}
\newcommand{\gy}{g_{_{\rm Y}}}
\newcommand{\alphaw}{\alpha_{_{\rm W}}}
\newcommand{\ncs}{N_{_{\rm CS}}} % Chern-Simons number
\newcommand{\nb}{n_{_{\rm B}}}
\newcommand{\be}{\begin{equation}}
\newcommand{\ee}{\end{equation}}
\newcommand{\ba}{\begin{array}}
\newcommand{\ea}{\end{array}}
\newcommand{\baa}{\begin{array}}
\newcommand{\eaa}{\end{array}}
\newcommand{\bea}{\begin{eqnarray}}
\newcommand{\eea}{\end{eqnarray}}
\newcommand{\half}{{1\over2}}
\newcommand{\trace}{{\rm Tr}}
\newcommand{\HiggN}{\varphi_0} %Normalized continuum Higgs field 0-mode 
\newcommand{\infN}{\infl_0} %Normalized continuum inflaton  field 0-mode 
\newcommand{\delw}{\delta_w}
\newcommand{\grad}{\vec\nabla}
\newcommand{\curl}{\vec\nabla\times}
\newcommand{\A}{\vec A}
\newcommand{\E}{\vec E}
\newcommand{\B}{\vec B}
\newcommand{\J}{\vec J}
\newcommand{\uno}{1\!{\rm l}}
\newcommand{\m}{\bar m}
\newcommand{\p}{{\bf p}}
\newcommand{\mbar}{\bar m}
\newcommand{\eabc}{\gw\epsilon_{abc}\,}
\newcommand{\phys}{\xi}
\newcommand{\pmin}{p_{\mbox{\tiny min}}}
\newcommand{\pmax}{p_{\mbox{\tiny max}}}
\newcommand{\bi}{{\rm Bi}}
\newcommand{\Real}{{\rm Re}}
\newcommand{\DNCS}{\Delta N_{\mbox{\tiny CS}}(t)}
\renewcommand{\topfraction}{0.8}

\preprint{FT-UAM-03/07, IFT-UAM/CSIC-03-13, hep-ph/0304285}
\title{Chern-Simons production during preheating in hybrid inflation models}
\author{Juan Garc{\'\i}a-Bellido, Margarita Garc\'\i a P\'erez  and Antonio Gonz\'alez-Arroyo}
\affiliation{Departamento de F\'\i sica Te\'orica \ C-XI, Universidad
Aut\'onoma de Madrid, Cantoblanco, 28049 Madrid, Spain\\
Instituto de F\'\i sica Te\'orica \ C-XVI, Universidad
Aut\'onoma de Madrid, Cantoblanco, 28049 Madrid, Spain}
\date{May 23, 2003}
\pacs{98.80.Cq \\
Preprint \ FT-UAM-03/07, IFT-UAM/CSIC-03-13, hep-ph/0304285}
\begin{abstract}
We study the onset of symmetry breaking after hybrid inflation in a model
having the field content of the SU(2) gauge-scalar sector of the standard
model, coupled to a singlet inflaton. This process is studied in
(3+1)-dimensions in a fully nonperturbative way with the help of lattice
techniques within the classical approximation. We focus on the role played
by gauge fields and, in particular, on the generation of Chern-Simons
number. Our results are shown to be insensitive to the various cutoffs
introduced in our numerical approach. The
spectra preserves a large hierarchy between long and short-wavelength modes
during the whole period of symmetry breaking and Chern-Simons generation,
confirming that the dynamics is driven by the low momentum sector of the
theory. We establish that the Chern-Simons production mechanism is
associated with local sphaleron-like structures. The corresponding sphaleron
rates are of order $10^{-5}\ m^4$, which, within certain scenarios of
electroweak baryogenesis and a (not unnaturally large) additional source of
CP violation, could explain the present baryon asymmetry of the Universe.
\end{abstract}

\maketitle

%%%%%%%%%%%%%%%%%%%%%%%%%%%%%%%%%%%%
%%%      SECTION INTRODUCTION
%%%%%%%%%%%%%%%%%%%%%%%%%%%%%%%%%%%%

\section{Introduction}

Everything we see in the Universe, from planets and stars, to galaxies
and clusters of galaxies, is made out of matter, so where did the
antimatter in the Universe go? Is this the result of an accident, a
contingency during the evolution of the Universe, or is it
an inevitable consequence of some asymmetry in the laws of nature?
Theorists tend to believe that the observed excess of matter over
antimatter, $\eta = (n_{\rm B}-n_{\bar{\rm B}})/n_\gamma \sim
6\times10^{-10}$, comes from tiny differences in their fundamental
interactions soon after the end of inflation. 

It has been known since Sakharov's work that there are three necessary
(but not sufficient) conditions for the baryon asymmetry of the
Universe to develop~\cite{sakharov}. First, we need interactions that
do not conserve baryon number B, otherwise no asymmetry could be
produced in the first place. Second, C and CP symmetry must be
violated, in order to differentiate between matter and antimatter,
otherwise B nonconserving interactions would produce baryons and
antibaryons at the same rate, thus maintaining zero net baryon number.
Third, these processes should occur out of thermal equilibrium,
otherwise the net baryon number cannot change in time.

The possibility that baryogenesis could have occurred at the
electroweak scale is very appealing~\cite{RS}. The Standard Model is
baryon symmetric at the classical level, but violates B at the quantum
level, through the chiral anomaly.  Electroweak (EW) interactions
violate C and CP through the irreducible phase in the
Cabibbo-Kobayashi-Maskawa (CKM) matrix, but the magnitude of the
violation is probably insufficient to account for the observed baryon
asymmetry~\cite{RS}.

One of the most appealing mechanisms for generating the baryon asymmetry
of the Universe makes use of the nonperturbative
baryon-number-violating sphaleron interactions present in the
electroweak model at high temperatures~\cite{KRS}.  The usual scenario
invokes a strongly first-order phase transition to drive the
primordial plasma out of equilibrium and set the stage for
baryogenesis.  This scenario presupposes that the Universe was in
thermal equilibrium before and after the electroweak phase transition,
and far from it during the phase transition. Although there is a
mounting evidence in support of the standard Big-Bang theory up to 
nucleosynthesis temperatures of ${\cal O}(1$~MeV), the assumption that
the Universe was in thermal equilibrium at earlier times is merely a
result of a (plausible) theoretical extrapolation. Furthermore, the
electroweak phase transition is certainly not first
order~\cite{Rummukainen}, given the present lower bound on the mass of
the Higgs boson~\cite{PDG}. In order to account for the observed baryon
asymmetry and to prevent the later baryon wash-out, a stronger
deviation from thermal equilibrium is required~\cite{leptogenesis}. 

According to recent studies of reheating after inflation, the Universe could
have undergone a period of ``preheating''~\cite{KLS}, during which only
certain modes are highly populated, and the Universe remains for some time
very far from thermal equilibrium~\cite{nonthermal}.  Recently, a new
mechanism for electroweak baryogenesis was proposed~\cite{GGKS,KT}, based on
the nonperturbative production of long-wavelength gauge and Higgs field
configurations via parametric resonance at the end of inflation. Such
mechanism occurs very far from equilibrium and could be very efficient in
producing the baryon asymmetry of the Universe at the EW scale. The very
nonequilibrium nature of preheating may facilitate the baryon number
generation, due to the sustained coherent oscillations of the scalar fields
involved~\cite{GBG,JGB,Dima,CGK}.

The proposed scenario suggests a picture of the early Universe in
which thermal equilibrium is maintained only up to temperatures of
${\cal O}(100$ GeV). The earlier history of the Universe is diluted by
a low-scale period of inflation, after which the Universe never
reheated above the electroweak scale. It is not easy to construct a
natural model of low-scale inflation. The main problem is to achieve
the extreme flatness of the effective potential in the inflaton
direction (i.e. the smallness of the inflaton mass in the false
vacuum) without fine-tuning~\cite{Lyth}. Although several models have been
proposed, the lack of naturalness remains a serious problem.  Perhaps
recent ideas related to large internal dimensions can provide a
solution~\cite{Creminelli}. We will not be concerned with such issues
here. The only qualitative feature of the low-energy inflation model
that is essential to us is that it provides a ``cold'' state as
initial condition for EW symmetry breaking. The main question to be
addressed here is whether electroweak baryogenesis can take place
under these circumstances.

As a specific model, a gauged Higgs-inflaton model of hybrid inflation
was considered in Ref.~\cite{GGKS}. In this model the inflaton is a
singlet coupled only to the Higgs boson, and induces dynamical electroweak
symmetry breaking as it slow-rolls below a critical value. The false
vacuum energy is quickly converted into a Higgs condensate with large
occupation numbers, in a process known as ``tachyonic preheating''
\cite{GBKLT}. The system evolves toward equilibrium while slowly
populating higher and higher momentum modes~\cite{TK}. The expansion
of the Universe at the electroweak scale is negligible compared to the
mass scales involved, so the energy density is conserved, and the
final reheating temperature $\trh$ is solely determined by the energy
stored initially in the inflaton field, that is, the false vacuum
energy density.

To make a quantitative test of the proposal, a (1+1)-dimensional
Abelian Higgs model~\cite{GRS,1p1} was studied in Ref.~\cite{GGKS}. It
contains all the relevant ingredients: an anomalous current which
relates the (1+1)-topological U(1) winding number to the global charge
(baryon number), plus a CP-violating operator dependent upon the Higgs
expectation value.  In Ref.~\cite{GGKS} the scenario was shown to be
very efficient in preventing the baryon wash-out, thanks to the
coherent oscillations of the Higgs condensate~\cite{GBG}.  Although
the results are very encouraging, they do not yet prove the validity
of the scenario since there might be specific signatures in
(3+1)-dimensions which prevent the production of baryons.  Other
authors have also studied this (1+1) dimensional
model~\cite{ST,ST2,CPR}, with similar conclusions regarding the
generation of baryon number asymmetry, but with different points of
view concerning its origin.

Since then, there have been several papers studying this scenario in
(3+1)-dimensions.  It was shown in Ref.~\cite{GBKLT} that preheating
does not proceed after hybrid inflation via parametric resonance, but
through spinodal instability, which induces the tachyonic growth of the
long-wavelength Higgs modes. The first studies were done in the quench
approximation~\cite{GBKLT,GBRM}, and later in the complete hybrid
inflation scenario with a time-dependent Higgs mass~\cite{JGB,ABC,CPR,GGG}.
These studies suggest that preheating in hybrid inflation models can be
very efficient in producing Higgs semiclassical modes at the electroweak
symmetry breaking. Most of these papers consider only the purely scalar
sector. However, in Ref.~\cite{CPR} an Abelian gauge 
field was introduced to study the topological defect formation.
Non-Abelian gauge fields with different initial conditions
were considered in Ref.~\cite{RSC} and in a set-up similar to ours in 
Refs.~\cite{ST2,SST}. 

The detailed way in which symmetry breaking occurs at the end of
hybrid inflation was analyzed with a (3+1)-lattice simulation in
Ref.~\cite{GGG}, from now on referred to as paper I. In that paper, we
considered the hybrid model in the absence of gauge fields.  We studied the
early stages of the dynamics of the Higgs modes, evolving from quantum
to classical behavior, and the subsequent nonperturbative evolution
through symmetry breaking.  This process is mediated by the formation
and growth of lumps, which later develop into bubbles (or spherical
shock waves) and their subsequent collision leads the way toward
thermalization.  However, it is expected that the gauge fields play
an important role in the process of electroweak symmetry breaking.
Furthermore, the relevance of gauge fields in our context is connected
to the generation of Chern-Simons (CS) number, which, via the chiral
anomaly, may induce baryon production in the presence of a
CP-violating interaction.

In the present paper we address a detailed study of the
post-inflation dynamics of the hybrid model in (3+1) dimensions,
including non-Abelian gauge fields.  
We will study the evolution of this model from the initial
conditions suggested by hybrid inflation, through the tachyonic
preheating stage on the way toward full thermalization.
Our main goal will be to  analyze the generation of Chern-Simons number in this
process. It is expected that after inclusion of a CP-violating
operator, this will generate the required baryon asymmetry.  However,
in this paper we are not including such a term, which we postpone for
future work. Another possible improvement to our model is the
inclusion of the full SU(2)$\times$U(1) gauge group. However, as generally
assumed, the simplified SU(2) model suffices to describe the sphaleron
transitions  and Chern-Simons number production~\cite{RS}.

From the methodological point of view, our paper clarifies the
importance of setting the right initial conditions of the non-Abelian
gauge fields for the study of baryogenesis at preheating after
inflation. In a recent paper~\cite{moore}, Guy Moore challenged the
viability of lattice methods for testing the scenario of
Ref.~\cite{GGKS}. He argued that the classical evolution of the gauge
fields produced spectra that were soon dominated by high-momentum
modes and thus prone to large lattice systematic errors. As we will
show, this is not the case when the appropriate set of initial
conditions for both the Higgs and gauge fields is chosen. The spectra
of all fields involved remain strongly hierarchical, with the
occupation of long-wave modes several orders of magnitude larger than
that of the dangerous high-momentum modes throughout its evolution at
symmetry breaking and beyond the region of interest for Chern-Simons
number production.  Moreover, our results are stable under changes of
the cutoffs introduced in the procedure, thus providing a consistent
framework for analyzing the real-time evolution of the
inflaton-Higgs-gauge model.

 The paper is organized as follows. In Section II we describe the hybrid
model which we will study. We also give a brief explanation of
our choice of parameters, and of our post-inflation initial
conditions. In Section III we describe the methodology used to study the
dynamics of this system through symmetry breaking.  First, we give a
short recollection of our results in paper I, which show the fast
growth of certain Higgs modes and their transition from quantum to
classical behavior. This stage sets the stochastic initial conditions
for the subsequent classical real time evolution of the full
inflaton-Higgs-gauge system.  Then, we describe our use of lattice
techniques to approach the latter problem, as well as the specific
modifications necessary to handle the initial conditions in the presence
of gauge fields.  To facilitate reading, we collect some of the
technical aspects in two appendices. In Appendix A we derive the
equations of motion on the lattice, while in Appendix B we give some
further technical information about our lattice implementation of the
Gauss constraint at the initial time.  Section III also
describes the cutoffs which our approximation introduces and the ranges
in which they allow physical results to be extracted.

In Section IV we present and make a detailed analysis of our
results. First, we display the behavior of spatial averages of scalar
fields and energy fractions. This information monitors the transition to
symmetry breaking and the relative importance of each type of degree of
freedom. It also helps us to test insensitivity to cutoffs and other
choices of our numerical implementation.  This is also connected to the
information on spectra which is presented next. Finally, we turn our
attention to our main goal: Chern-Simons number production. We show
compelling evidence of the physical character of this generation and
investigate its connection to the appearance of local structures. 
In Section~V we  discuss  how these results can be connected to the
generation of baryon number. A thorough investigation of the latter
aspect lies out of the scope of this paper and is, therefore, deferred
for future works.  A short summary of the conclusions is presented in
Section~VI.

%%%%%%%%%%%%%%%%%%%%%%%%%%%%%%%%%%%%
%%%      SECTION MODEL
%%%%%%%%%%%%%%%%%%%%%%%%%%%%%%%%%%%%

\section{The model}
\label{model}

The hybrid model we are considering is a simple generalization of the
Standard Model symmetry breaking sector. The Lagrangian comprises five
scalar fields, a singlet inflaton, $\infl$, and a Higgs SU(2)-doublet,
$\Phi=\half(\phi_0\,1\!{\rm l}+i\phi^a\tau_a)$, where $\vec{\tau}$ are the
Pauli matrices, together with an SU(2) gauge field, $A_\mu^a$:
\begin{equation}\nonumber
{\cal L} = - {1\over4}F^a_{\mu\nu}F^{\mu\nu}_a + 
{\rm Tr}((D_\mu\Phi)^\dag D^\mu\Phi) + \half (\partial_\mu\infl)^2 - V(\Phi,\infl)\,, 
\end{equation}
where the covariant derivative is $D_\mu = \partial_\mu - {i\over2}
\gw A_\mu^a\tau_a$, with $\gw$ the SU(2) gauge coupling.
The expression  of the gauge field strength is given by
\begin{equation}
F^a_{\mu\nu} = \partial_\mu A_\nu^a - \partial_\nu A_\mu^a +
\gw \epsilon^{abc} A_\mu^b A_\nu^c\,.
\end{equation}

The scalar potential has the usual Higgs term
plus a coupling to a massive inflaton. Taking  
$ {\rm Tr}\, \Phi^\dag\Phi =
\half(\phi_0^2 + \phi^a\phi_a)\equiv\half|\phi|^2$ we have
\bea
&V(\Phi,\infl) = \frac{\lambda}{4}\left(|\phi|^2 - v^2\right)^2 +
\frac{g^2}{2}\infl^2|\phi|^2 + \half \mu^2 \infl^2 \\
&\hspace{.5cm}=V_0 + \half(g^2\infl^2-m^2)\,|\phi|^2 + \frac{\lambda}{4}
|\phi|^4 + \half \mu^2 \infl^2\nonumber \,,
\eea
where $\mu$ is the mass of the inflaton in the false vacuum, and
$m\equiv\sqrt\lambda\,v$; $\,v=246$ GeV is the expectation value of
the Higgs boson in the true vacuum.  The Higgs mass in the true vacuum is
determined by its self-coupling: $\mh \equiv \sqrt{2\lambda}\,v$,
while the mass of the inflaton in the true vacuum is given by 
$m_{\rm I} \equiv gv\gg\mu$.

Our aim is to study the evolution of this system from the end of
inflation to thermalization, starting with the initial conditions
described below.

\subsection{Initial conditions}

It is the effective false vacuum energy $V_0 = {1\over4}\lambda v^4 +
\half \mu^2 \infl_0^2$ that drives the (relatively short) period of
hybrid inflation, during which the inflaton field is described by its
homogeneous mode $\chi_0 \equiv \langle\chi\rangle$. Inflation ends when
this mode slow-rolls below the bifurcation point $t=t_c$
($\chi_0(t_c)=\chi_c\equiv m/g$). Around this time the inflaton behaves as
$$\chi_0 = \chi_c\,(1 - V\,m(t-t_c)),$$
where $V$ is the dimensionless
velocity of the inflaton (defined by this equation).  The actual value
of $V$ depends very much on the model and the scale of inflation, and
we will treat it here as an arbitrary model parameter.  At the
bifurcation point, the Higgs boson is massless, $m^2_\phi =
m^2\,(\chi_0^2/\chi_c^2 - 1) \approx - 2V\,m^3 (t-t_c)$. After this
point, the Higgs field acquires a negative time-dependent mass-squared
and long-wavelength modes will grow exponentially, driving the process
of symmetry breaking~\cite{GBKLT}. Typically the speed of the inflaton
is such that the process takes place in less than one Hubble time, a
condition known as the ``waterfall'' condition~\cite{hybrid,GBL},
which ensures the absence of a second period of inflation after the
bifurcation point~\cite{GBLW}.

Since inflation dilutes any previous fluctuations and/or particles, the
Universe is empty and cold at the end of this period. Therefore, we will
assume that both the Higgs and the gauge fields are in the de Sitter
vacuum. In fact, since for electroweak-scale inflation the rate of
expansion is negligible compared to any other scale, $H\approx 10^{-5}$
eV $\ll v$, the de Sitter vacuum state is equivalent to the Minkowski
vacuum for the range of momenta we are considering.

\subsection{Model parameters}

The model has a handful of parameters, from couplings between
different fields to self-couplings and masses. Although a wide range
of parameters can in principle be chosen for the model, specially in
the inflaton-Higgs sector (since it has never been measured), we will
choose particular values for definiteness. For example, the
Higgs-inflaton coupling $g$ will be chosen to be specifically related
to the Higgs boson self-coupling $\lambda$ (which determines its true
vacuum mass) as $g^2 = 2\lambda$.  As described in paper I, such a
choice, suggested by some supersymmetric versions of hybrid
inflation~\cite{BGK}, significantly simplifies the description of the
evolution of the Higgs and inflaton fields after symmetry breaking and
for this reason we have retained such a relationship in this paper.
On the other hand, the value of $\lambda$ or $\gw$ determines the
Higgs and $W$ boson masses, respectively, so we cannot chose them
arbitrarily. However, in this paper our main goal is to illustrate the
dynamics itself, rather than matching the constraints given by
experiment. Thus, we have chosen a range of values of the parameters,
given in Table~\ref{models}. Models A1-A4 have different values for
the ratio of masses $\mh/\mw$, while this ratio is the same for models
B and A1.  As we will describe later, in selecting the parameters it
is important to take into account the sensitivity of the results to
cutoff effects. Most of our results are obtained for model A1, for
which this sensitivity is supposed to be smaller.  In addition, and in
order to improve our approximations, a relatively large value of $V$,
equal to $0.024$, has been taken for all our models. This allows, for
reasons that will also be described later, to increase the value of
the ultraviolet cutoff. Note that large values of $V$ are expected in
hybrid models where the flat direction of the inflaton is tilted by
radiative corrections~\cite{DSS}.

\begin{table}
\begin{tabular}{||c||c|c|c||}\hline\hline
Model & $\lambda=g^2/2$  & $\gw$ & $\mh/\mw$ \\ \hline \hline
 A1 & $0.00675$ & $0.05$ & $4.65$ \\ \hline
 A2 & $0.00675$ & $0.0825$ & $2.82$ \\ \hline
 A3 & $0.00675$ & $0.1$ & $2.32$ \\ \hline
 A4 & $0.00675$ & $0.15$ & $1.55$ \\ \hline
 B & $0.2885$ & $0.3269$ & $4.65$ \\ \hline
\end{tabular}
\caption{List of model parameters used in our analysis. 
For all of them we have taken the inflaton velocity $V=0.024$ and the 
inflaton bare mass $\mu = 10^{-5} g v \approx 0$.}
\label{models}
\end{table}

%%%%%%%%%%%%%%%%%%%%%%%%%%%%%%%%%%%%
%%%%%%    SECTION METHODOLOGY
%%%%%%%%%%%%%%%%%%%%%%%%%%%%%%%%%%%%

\section{Methodology}\label{Methodology}

In this Section we discuss our approach to the study of symmetry
breaking. We first recall how in the initial stages of the
evolution of the system, the long-wavelength modes evolve from quantum
to classical behavior. This justifies our main approximation. Then we
describe our lattice approximation to the classical equations of
motion, as well as the determination of the initial conditions for the
classical evolution.

%%%%%%%%%  subsection classical 

\subsection{Transition to classical behavior}
\label{classical}

The problem of determining the time evolution of a quantum field
theory is outstandingly difficult. Fortunately, in some cases some
analytic control is possible because only a few degrees of freedom are
relevant, or else because perturbative techniques are applicable. Our
particular problem, however, is both nonlinear and nonperturbative
and involves many degrees of freedom. Moreover, the presence of gauge
fields just complicates matters further.

A first-principles approach to nonperturbative quantum field theory is
provided by the lattice formulation, in which the gauge principle is
easily incorporated~\cite{wilson}. The existing powerful lattice field
theory numerical methods rest on the path integral formulation in
Euclidean space and the existence of a probability measure in field
space.  However, the problem in which we are interested in is a
dynamical process far from equilibrium, and the corresponding Minkowski
path integral formulation is neither mathematically well founded, nor
appropriate for numerical studies. 

There are a series of alternative nonperturbative methods which
different research groups have used to obtain physical results in
situations similar to ours. These include Hartree's
approximations~\cite{CHKMPA} to go beyond perturbation theory or large
N techniques~\cite{BdV,BHP}. It is, no doubt, desirable to approach this
and similar problems with all available tools. 

In the present paper we will use an alternative approach: the classical
approximation.  It consists of substituting the quantum evolution of the
system, dominated initially by the long-wave semiclassical modes,
by its classical evolution, for which there are
feasible numerical methods available.  The quantum nature of the problem
remains in the stochastic character of the initial conditions.  The
advantage of the method is that it is fully nonlinear and
nonperturbative, allows the use of gauge fields and gives access to the
quantities we are interested in.  This approximation has been used
previously by several authors in various contexts within the standard
cosmological literature~\cite{PS}. It has also been applied to the study
of preheating after inflation~\cite{TK,PR,FT,GBKLT}.  In paper
I we built on work of previous authors and gave a detailed
justification of the validity of this approximation for our particular
situation and in the absence of gauge fields.  The main aspects of the
method are the following.

We start the evolution of the system at the critical time $t_c$, at
which the effective mass of the Higgs boson vanishes, putting all the
modes in their (free field) Minkowski ground states. Strictly speaking
the de-Sitter vacuum is more appropriate, but the difference turns out
to be negligible for the relevant modes, due to the minute rate of
expansion. Initially, since the quantum fluctuations are not large,
and whenever the couplings are small, the nonlinear terms in the
Hamiltonian of the system can be neglected. Then the quantum evolution
is Gaussian and can be studied {\em exactly}. The Hamiltonian for the
Higgs modes inherits a time dependence through the coupling to the
time-dependent inflaton homogeneous mode. This time dependence can
always be taken to be linear for a sufficiently short time interval,
as was done in Section~\ref{model}.  Most of the Higgs or inflaton
modes evolve in a characteristic harmonic oscillator fashion with a
frequency depending on the mode in question, except for the case of
the low-frequency modes of the Higgs field, which become tachyonic and
grow exponentially. By looking at the vacuum expectation values of
products of these fields at later times, one realizes that after a
while these modes behave and evolve like classical modes. The process
is very fast and therefore the remaining harmonic modes can be assumed
to remain in their initial quantum vacuum (ground) state.
 
The fast growth in size of the Higgs field expectation value boosts
the nonlinear terms and eventually drives the system into a state
where the nonlinear dynamics, including the backreaction to the
inflaton field, are crucial. For the whole approximation to be useful,
this must happen after the time in which the low-frequency Higgs modes
begin evolving as classical fields.  Thus, the main philosophy
underlying our method is to turn on classical evolution at a time
$t_i>t_c$, where low momentum modes are already classical, while
nonlinearities (including backreaction effects) still remain small.
In paper I, we showed that, indeed, there is a time interval for $t_i$
in which these conditions hold.  We tested that our results were
insensitive to the particular value of $t_i$ provided it lies within
this window. This is so despite the large differences in size of the
initial Higgs modes corresponding to different choices.  Similar ideas
have also been used by Smit {\em et al.}~\cite{SVS,ST} for the case of
a quench. Our approach differs from that of other authors who use the
classical evolution starting from $t=t_c$~\cite{GBKLT,CPR}.
 
There is a relatively sharp separation in momenta between classical
modes and quantum modes. The latter, being nontachyonic, have evolved
only slightly from their initial vacuum state. A distinctive feature
of quantum field theory is that high-momentum modes produce a sizable
(divergent) contribution to some observables. However, most of this
effect goes into a renormalization of the parameters of the theory.
Thus, our approach, which is similar to that of Smit et
al.~\cite{SVS,ST}, is to cutoff the modes that have not yet entered
the classical regime at $t_i$, and use a classical theory with
renormalized couplings~\cite{GGG} thereafter. We find that our results
are robust to changes in the way the cutoff is implemented.

In the present paper we follow the same strategy as in paper I, but in the
presence of non-Abelian gauge fields. The initial quantum evolution of
gauge fields is also relatively slow, since there are no tachyonic
modes. Therefore, it is assumed not to affect substantially the initial
conditions of the classical system.  The only limitation to these idea
is determined by Gauss law, which relates the Coulomb (electric)
component of the gauge field to the charge distribution produced by the
classical Higgs field. All the remaining (propagating) modes of the
gauge field are taken to be zero.  Choosing nonzero but small values
does not affect our results substantially provided we cut them off at
large momenta.  This seems consistent with the philosophy applied to the
Higgs field.  We believe, however, that the dynamics of gauge fields can
play a crucial role at later times, and in particular in the mechanism
of Chern-Simons production. This paper is devoted to investigating this
point.

%%%%%%%%%  subsection lattice

\subsection{Lattice approximation to the classical evolution}
\label{lattice}

In the previous Subsection we have explained the main idea underlying
our approach to the problem. We study the nonlinear and
nonperturbative classical evolution of the system starting at time
$t=t_i>t_c$, with stochastic initial conditions which match the
expectation values obtained at $t_i$ from the quantum linear evolution
of the Higgs system~\cite{GGG}. 
The classical equations of motion are
\begin{eqnarray}
D_\mu D^\mu\Phi + \lambda\left(|\phi|^2 - v^2\right)^2 \Phi
+ g^2\infl^2 \Phi&=& 0\,,\nonumber \\
\label{class_eqs}
\partial_\mu\partial^\mu\infl + \mu^2\infl +
g^2|\phi|^2\, \infl&=& 0 \,, \\
D^\nu F^a_{\mu\nu} - j^a_\mu &=& 0 \,,\nonumber 
\end{eqnarray}
where 
\be\label{gicur}
j^a_\mu = i\gw {\rm Tr}\,[\tau^a (\Phi^\dag D_\mu \Phi - (D_\mu
\Phi)^\dag\Phi)]
\ee
is the current induced by the charged Higgs field.

The full nonlinear evolution of the system can be studied using
lattice techniques. Our approach is standard. We discretize the
classical equations of motion in both space and time, but preserving
the gauge invariance of the system \cite{Ambjorn}. Full details are given in
Appendix A. The timelike lattice spacing $a_t$ must be smaller than
the spatial one $a_s\equiv a$ for the stability of the discretized equations.

In addition to the ultraviolet cutoff, $a$, one must introduce an
infrared cutoff by putting the system in a box with periodic boundary
conditions. We have studied $32^3$, $48^3$, $64^3$ lattices for large
statistics studies, and $80^3$ for a few configurations.  Computer
memory and CPU resources limit us from reaching much bigger
lattices. One of our main goals has been to check that the physical
results are not affected by the presence of these cutoffs and other
details of our practical method, as for example the choice of $t_i$. For
that to happen these cutoffs have to be taken within appropriate limits
fixed by the relevant physical scales involved. In our problem there are
several scales depending on the parameters of the model and controlling
different time regimes and observables. Thus, it is not always an easy
matter to place these scales in the window defined by our ultraviolet
and infrared cutoffs. Since in this paper we are more interested in
understanding the phenomena themselves rather than in extracting
specific phenomenological predictions, our attitude has been to modify
the parameters of the model to place ourselves in a region where our
results are more insensitive to the cutoffs.  This is, no doubt, a
necessary first step to determine the requirements and viability of the
study of any particular model.

In determining a good set of parameters, there are a few essential
ingredients.  First, the relevant set of initial momenta at time $t_i$
has to be well contained within the range $[\pmin,\pmax]$, where
$\pmax=\pi/a$ is the UV cutoff and where momentum is quantized in
units of $\pmin=2\pi/L$.  For the typical values of $t_i$, relevant
momenta are $k \lsim \sqrt 2 M$, where $M= (2V)^{1/3} m$ is a
characteristic scale associated with the inflaton initial velocity which
also enters in the determination of the bubble sizes and collisions
\cite{GGG}.  In this paper we have chosen a much higher inflaton
velocity $V$ than the one in Paper I. This reduces the hierarchy of
scales $M/m$ from 0.18 to 0.36 and thus allows to reduce the typical
lattice spacings by a factor of two. The typical values of the lattice
spacing in our simulations thus range from $ma=0.65$ to
$ma=1.3$. Another consideration concerns the time scales for
backreaction ($t_{br}$) and symmetry breaking ($t_{sb}$).  There should
be a hierarchy of scales such that $t_i \ll t_{br} \ll t_{sb}$. For
this to happen, neither $\lambda$ nor $mVt_i$ can be too large. Finally
there is the ratio between the Higgs and W boson masses, which has
to be appropriately chosen in order to avoid scaling violations in the
fluctuations of the Chern-Simons number, as will be discussed in detail
in Section \ref{cherns}.

A particularly good choice of parameters determines what we call model
A1 (see table I). This has been studied extensively and, as we will see
later, the insensitivity to cutoff effects is quite satisfactory for
it.  For this model $\gw = 0.05$, and the Higgs to W boson mass ratio is
given by $\mh/\mw=4.65$. We have then modified the value of the gauge
coupling constant $\gw$ (models A2-4) to cover a range of values for this
ratio down to $\mh/\mw\sim 1$. As we decrease the ratio our results, in
particular those concerning the Chern-Simons number, become increasingly
sensitive to the cutoffs. This is related to the fact that the relevant
scale for Chern-Simons generation is $\mw a$ which increases, for fixed 
$\mh a \sim 1$, when decreasing the $\mh/\mw$ mass ratio.   
We have also studied a situation
(model B) with the same $\mh/\mw$ ratio as in model A1, but with a value
of the gauge coupling constant $\gw=0.3269$ closer to that of the
standard model. This model might be subject to stronger systematic errors.
The Higgs boson self-coupling is, in this case, relatively
large and one might worry about the size of the initial fluctuations
being too large to ignore the backreaction of the inflaton and the
nonlinearities in the initial conditions.

\subsubsection{Initial conditions for classical evolution}
\label{initial}

%%%%%%%%%  subsection initial

In this Subsection we specify in greater detail the initial conditions
for the evolution of the classical system. For the Higgs and inflaton
fields they are fixed in the same fashion as in paper I, which is
summarized in what follows.  The inflaton field is given by the
homogeneous component alone, with a value given by $\chi_0(t_i)=\chi_c
(1 - V m (t_i-t_c))$ and conjugate momentum $\dot \chi_0(t_i)=-\chi_c
V m$. Higgs field modes with $ k/M > \sqrt{M t_i}\ $ are set to zero,
while the remaining ones are determined by a Gaussian random field of
zeromean and amplitude distributed according to the Rayleigh
distribution:
\be\label{rayleigh}
P(|\phi_k|)\,d|\phi_k|\,d\theta_k =
\exp\Big(\!-{|\phi_k|^2\over\sigma_k^2}\Big)\,
{d|\phi_k|^2\over\sigma_k^2}\,{d\theta_k\over2\pi}\,,
\ee
with a uniform random phase $\theta_k \in [0,2\pi]$ and dispersion given
by $\sigma_k^2 = k^{-3} P(k,t_i)$. $P(k,t_i)$ is the power spectrum of
the initial Higgs boson quantum fluctuations, in the background of the
homogeneous inflaton, computed in the linear approximation.  It is
defined through: 

\be
\langle|\phi (\vec x, t_i)|^2\rangle \equiv   
\sum_{\vec k\ne \vec 0}  {1 \over |\vec k|} \,P(\vec k,t_i) 
\equiv  
\sum_{\vec k\ne \vec 0} |f_k(t_i)|^2\,,
\ee
In the region of low momentum modes, $P(\vec k,t_i)$ is very well described 
by
\be\label{Papp}
P_{\rm app}(\vec k,t_i) = A(Mt_i)\,k^2\,e^{-B(Mt_i)\,k^2}\,,
\ee
where $A(Mt_i)$ and $B(Mt_i)$ are parameters extracted from a fit to
this form of the exact power spectrum given in Paper I.
In the classical limit the conjugate momentum $\dot \phi_k(t_i)$ is 
uniquely determined through $\dot \phi_k(t_i)  = C(k,t_i) \  \phi_k(t_i)$ with 
$C(k,t_i)= \Real (f_k(t_i)\dot f_k^*(t_i))/ |f_k(t_i)|^2 $. We have used a
fit to the latter quantity in our numerical implementation of the initial
conditions. 

Finally, we focus on the initial conditions for the gauge fields.
Since there are no tachyonic gauge field modes the strategy applied to
the Higgs-inflaton system suggests setting all initial gauge modes to
zero.  Here, however, we have to take into consideration gauge
invariance. As long as we focus on the time evolution of gauge
invariant quantities, there is no difference in choosing one gauge or
another. In the Hamiltonian formulation the most appropriate one is
the $A_0=0$ gauge. In this gauge we fix time-dependent gauge
transformations. The dynamical fields are then the spatial components
$A_i$, which satisfy standard Euler-Lagrange equations of motion
following from the Lagrangian. One is still free to perform
time-independent gauge transformations, which are a symmetry of the
Hamiltonian.  Although, $A_0$ is not a dynamical degree of freedom,
the associated field equation is still present: the non-Abelian
generalization of the Gauss law.  It takes the form of a constraint,
which has to be imposed on the initial condition, and relates the
gauge field to the classical current generated by the Higgs field. The
remaining field equations ensure that this constraint will continue to
be satisfied for all future times.  All this scheme translates to the
lattice formulation, as shown in Appendix A.  Thus our discretized
equations of motion guarantee that if the lattice version of the Gauss
constraint is imposed initially, it will continue to hold at later
times.  As in Refs.~\cite{GRS,1p1} there are slight violations which
are introduced by rounding errors of the numerical procedure. These,
nevertheless, remain very small for the range of times which we
consider in this paper.

In conclusion, the necessity of satisfying the Gauss constraint without
modifying the initial distribution of the Higgs boson forces the
initial gauge field to be nonzero. Our criterion has been to
introduce only those longitudinal components of the gauge potential
necessary to satisfy the Gauss constraint. The procedure that we have
used in practice is very similar to the one adopted in Ref.~\cite{ST}
for Abelian gauge fields and will be described in detail in Appendix
B. There we also show that the results are insensitive to details of
the implementation.

\begin{figure}[htb]
\vspace*{8.5cm}
\includegraphics{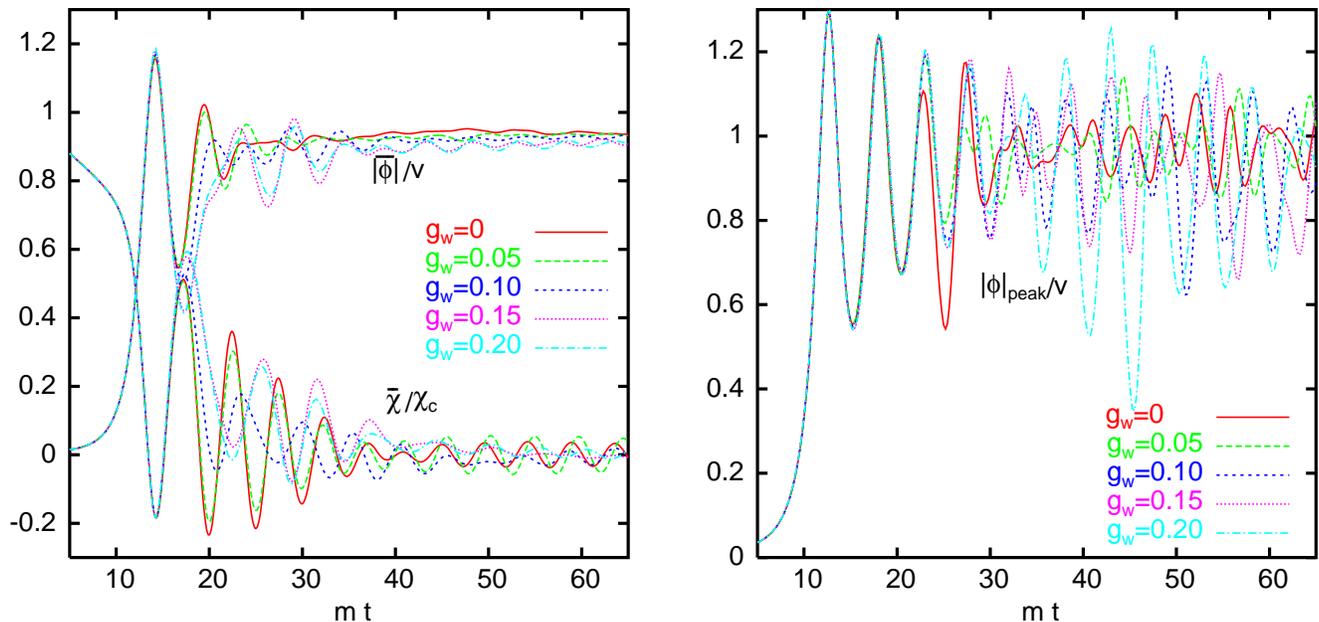}
\caption{ Left: The time evolution of the spatial averages of the normalized
Higgs field ($\overline{
|\phi|}/v$) and inflaton
($\overline{\chi}/\chi_c $) for a single configuration of model A
($\lambda=g^2/2=0.00675$, $V=0.024$) and various values of the
gauge coupling $\gw$ (see Table~\ref{models}). Right: The time evolution
of the top of the highest lump  $ |\phi|_{\rm peak}/v$.}
\label{means.conf}
\end{figure}

%%%%%%%%%%%%%%%%%%%%%%%%%%%%%%%%%%%%
%%%%%%    SECTION RESULTS
%%%%%%%%%%%%%%%%%%%%%%%%%%%%%%%%%%%%

\section{Results}

In this Section we present the main results of our investigation. 
First we describe global quantities as  average field values 
and energy fractions. Next we analyze the gauge invariant spectra 
of the system. Finally, we focus on the Chern-Simons number production
and investigate its relation to local space-time structures. 
For all quantities we test the sensitivity of the results 
to  the lattice spacing and  volume.

\begin{figure}[htb]
\vspace*{12cm}
\includegraphics{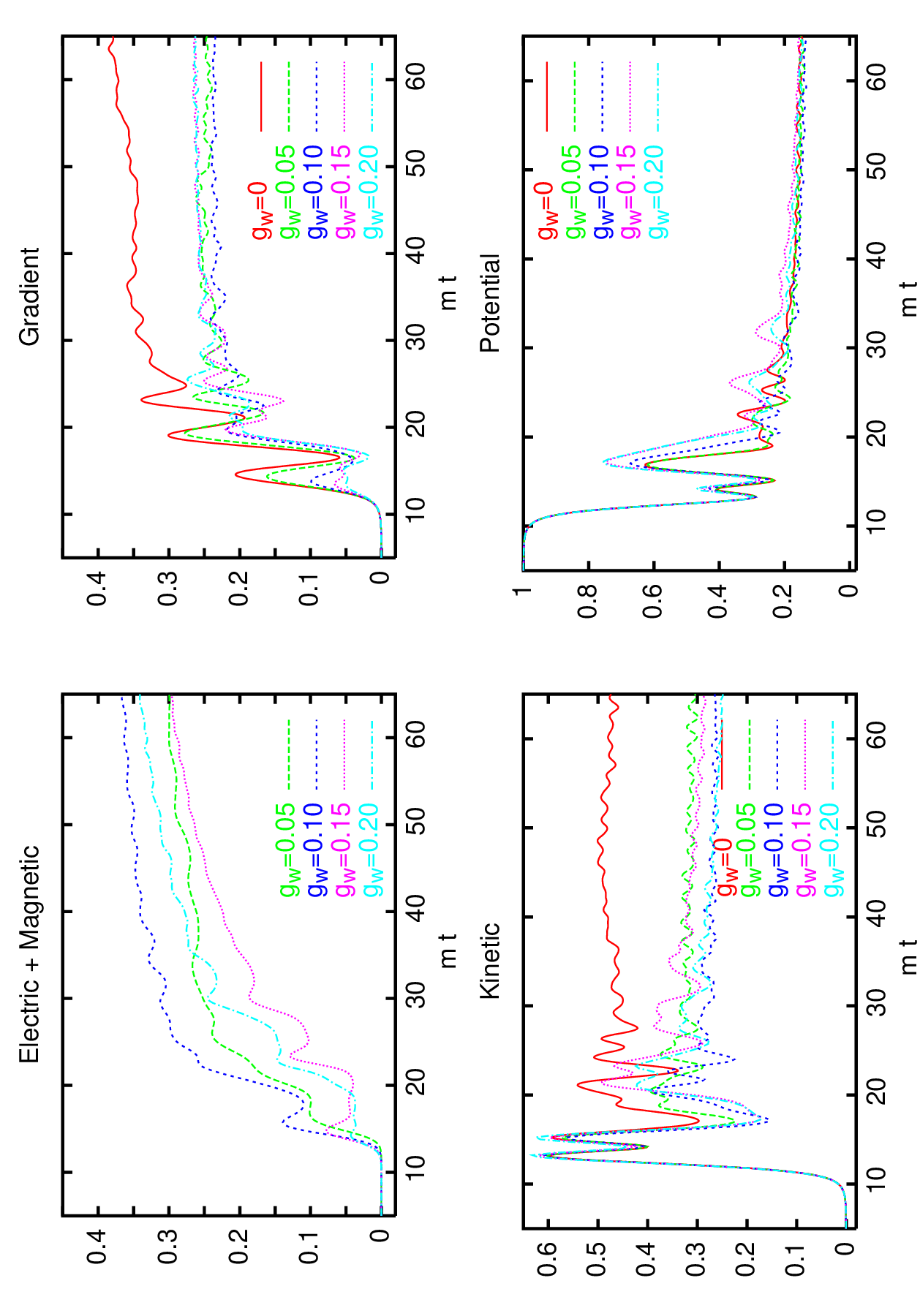}
\caption{ The time evolution of the  energy fractions
for a single configuration of model A and
various values of the gauge coupling $\gw$. }
\label{ener.conf}
\end{figure}

\begin{figure}[htb]
\vspace*{14.5cm}
\includegraphics{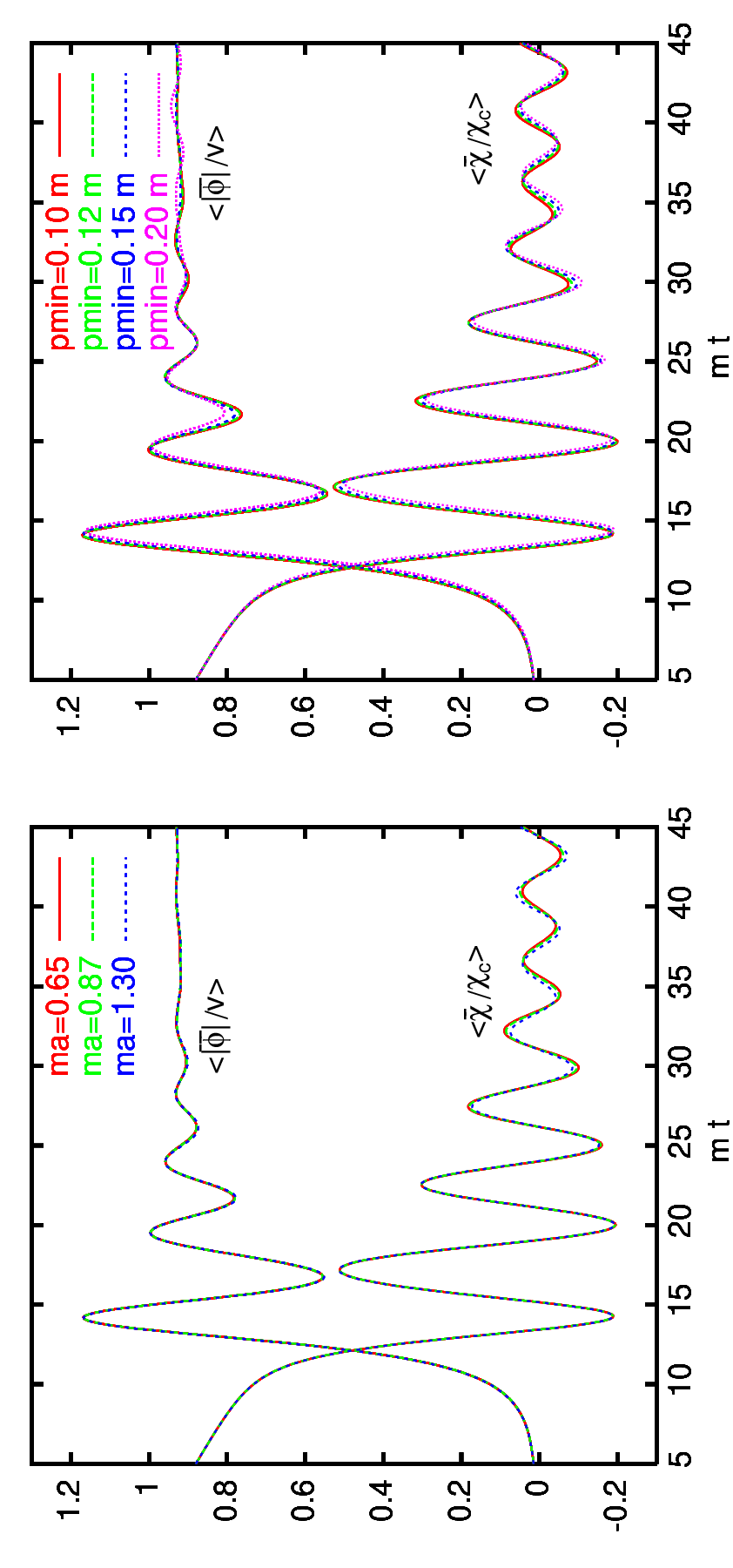}
\includegraphics{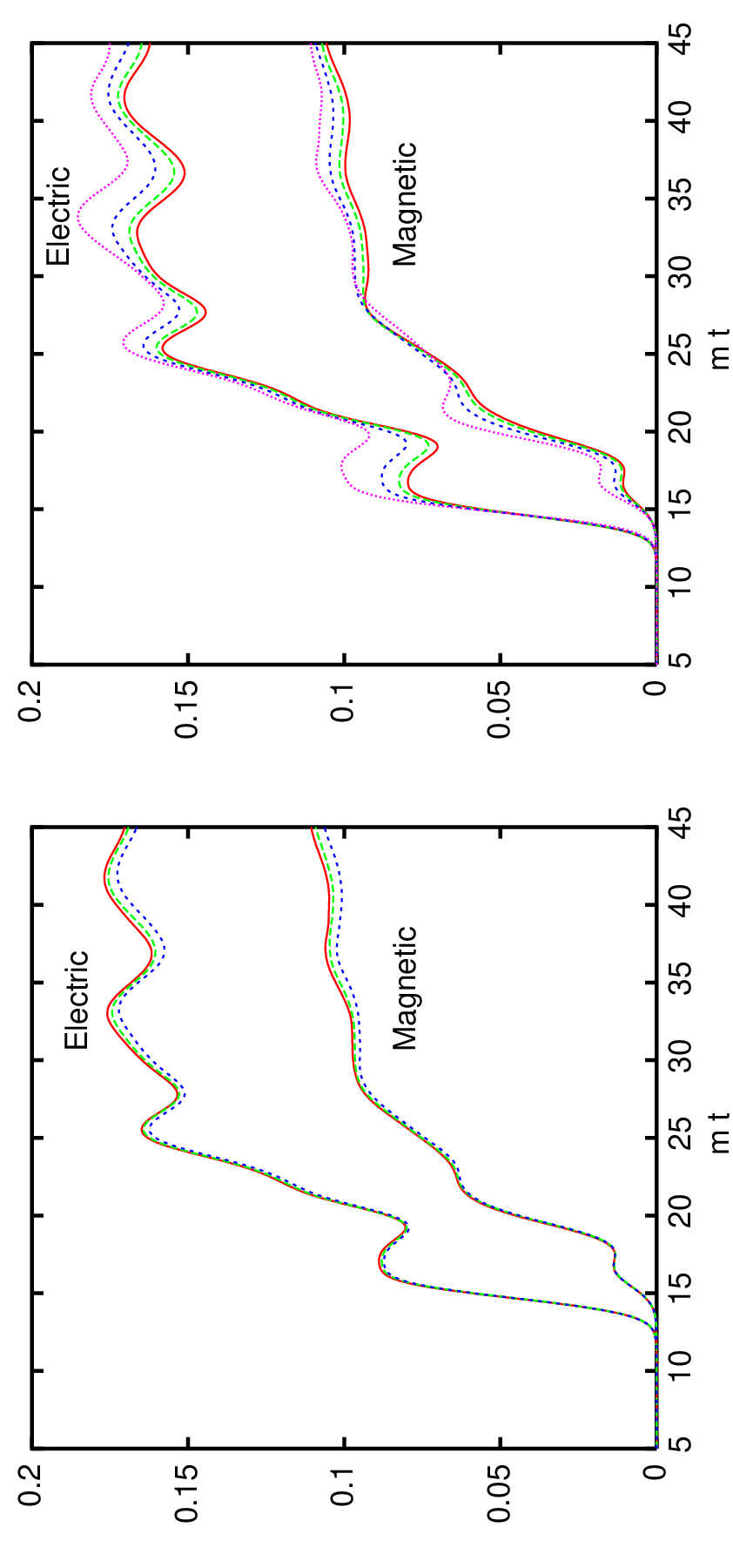}
\caption{Left: The UV cutoff dependence of the normalized $\langle
\overline{|\phi|}\rangle$ and $\langle \overline{\chi} \rangle$ and
average electric and magnetic energy fractions.  Right: The volume
dependence of $\langle \overline{|\phi|}\rangle$ and $\langle
\overline{\chi}\rangle$ and average electric and magnetic energy
fractions.  Parameters as in model A1 in Table \ref{models}.  }
\label{means}
\end{figure}

\begin{figure}[htb]
\vspace*{8.5cm}
\includegraphics{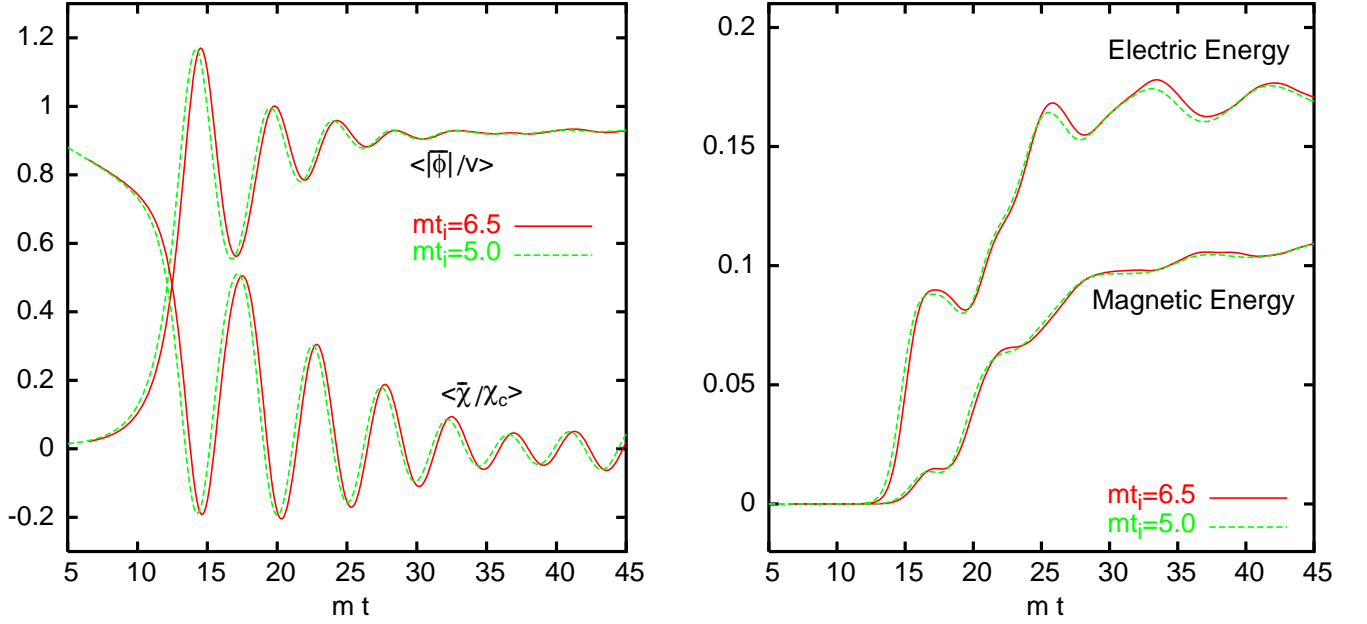}
\caption{The time evolution of $\langle \overline{|\phi|}/v\rangle$,
$\langle \overline{\chi}/\chi_c\rangle$ and gauge energy fractions for
two different initial times  $mt_i=5$ and $mt_i=6.5$ (model A1).
}
\label{means_tidep}
\end{figure}

\begin{figure}[htb]
\vspace*{8.cm}
\includegraphics{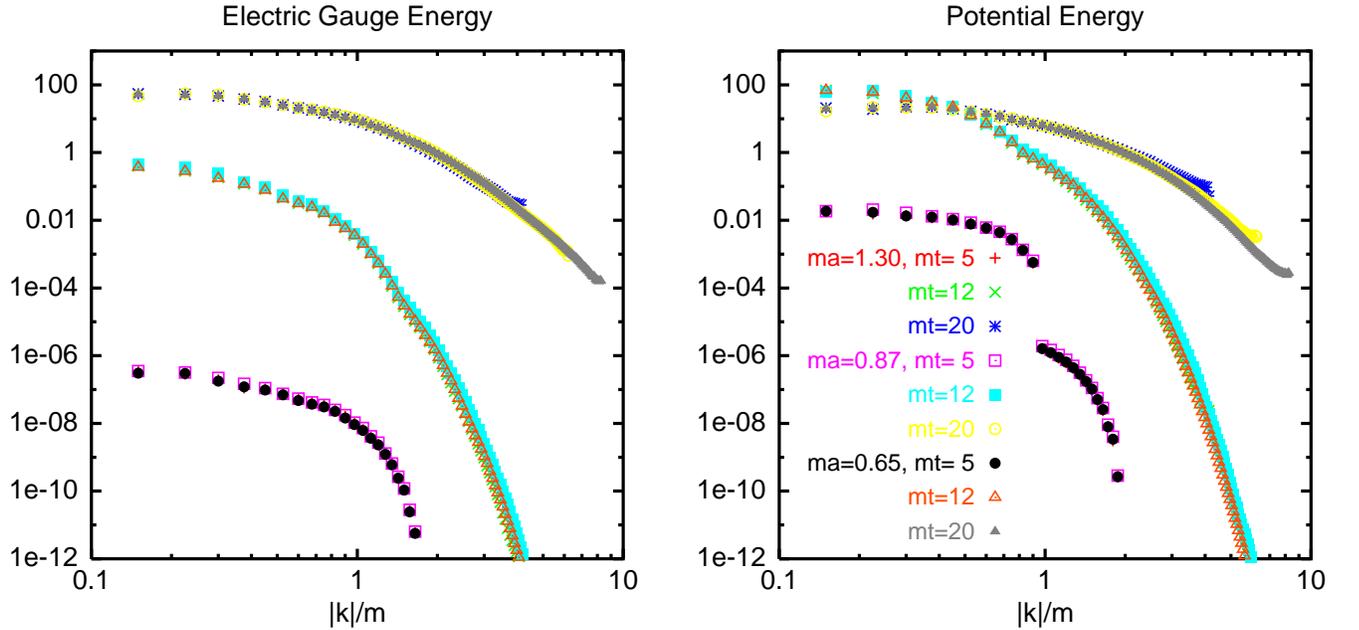}
\caption{Dependence on the UV cutoff of the
  Fourier spectra of the (pure gauge) electric energy and
  Higgs-inflaton contribution to the potential energy evaluated at
  $mt=5$, 12 and 20.  Results are averaged over 10 configurations with
  $\lambda=g^2/2=0.00675$, $V=0.024$, and $\gw=0.05$ (model A1), and
  $mt_i=5$. }
\label{spectra}
\end{figure}
\begin{figure}[htb]
\vspace*{8.cm}
\includegraphics{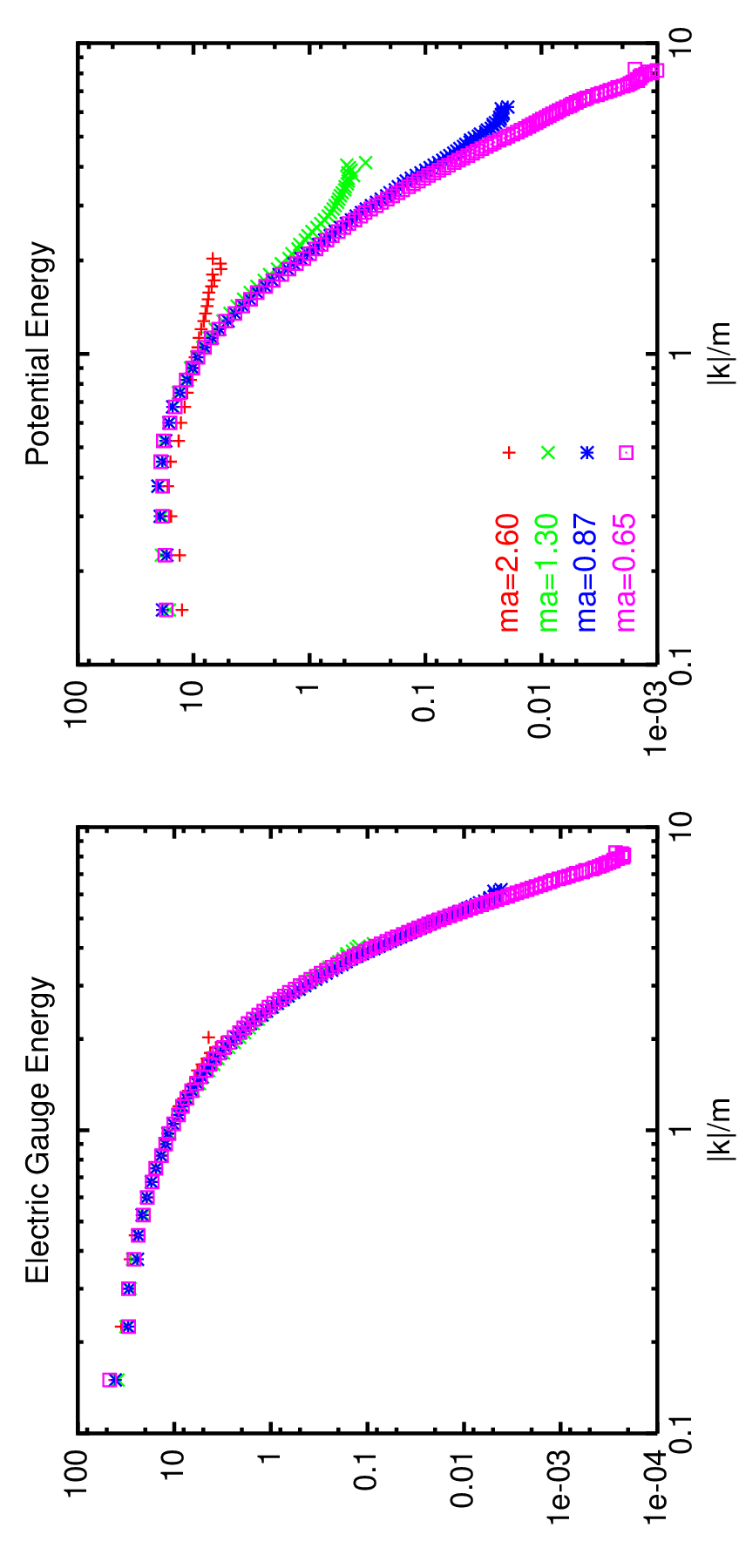}
\caption{Dependence on the UV cutoff of the Fourier spectra of the
electric and Higgs-inflaton potential parts of the energy for $mt=45$.
Parameters as in Fig. \ref{spectra}.}
\label{spectra_t45}
\end{figure}

%%%%%%%%%%%%%%%  subsection evolution
\subsection{Evolution of average  values}

The main stages of the dynamical evolution of the system leading to
the onset of symmetry breaking have been described in paper I, in the
absence of gauge fields.  This is a very inhomogeneous process that
proceeds via the formation of lumps in space that grow with time. The
location of the lumps and their initial height depends on the
particular realization of the Gaussian random initial conditions, but
the main features of the dynamics do not.  Once the Higgs field at the
top of the lump increases beyond the classical minimum of its
potential, it bounces and starts to oscillate.  In this period of
time, the locus of the local maxima of the Higgs field value has the
shape of a spherical bubble which grows with time.  Later on,
neighboring bubbles start to collide, producing higher-frequency
modes, which end up destroying the coherent oscillatory behavior on
the way toward thermalization of the system. A thorough description of
the details of this process has been presented in paper I for the
scalar sector of the model; here we will mostly concentrate on the
differences that are introduced by the coupling to the gauge fields.

A first category of observables monitoring this process is given by
the time evolution of the gauge invariant, spatial averages of
$|\phi(\vec{x},t)|$ and $\chi(\vec{x},t) $.  In Fig.~\ref{means.conf}
(left) we show an illustration of such a time evolution for model A from
the initial distribution at $mt_i=5$ up to $mt=65$, where the system
has stabilized around the true vacuum: the Higgs field gets close to
its vacuum expectation value and the inflaton oscillates around zero
with an amplitude that gets damped as time evolves.  As mentioned
previously the initial phases of this evolution involve the presence
of Higgs boson lumps in space that grow with time. Figure
\ref{means.conf} (right) shows the time evolution of the center of one
of those lumps. Both the spatial averages and the value at the top of
the lump are computed and shown for various values of the gauge
coupling, including zero coupling. From the comparison it is clear
that initially the dynamics is completely driven by the scalar sector
of the model; only at later times, once the Higgs and inflaton fields
have started oscillating around their vacuum expectation values, does
the presence of gauge fields start to affect the evolution of these
observables.

The moment in which the gauge fields start to play an essential role in
the evolution is more clearly signaled by the time evolution of the
total energy fractions. These quantities represent the relative contributions of
the pure gauge electric and magnetic terms in the Hamiltonian, kinetic,
gradient and potential terms of the Higgs and inflaton fields to 
the conserved total energy of the system.  In Fig.~\ref{ener.conf} we
display the electric plus magnetic energy fractions as a function of
time for a given configuration in model A.  The fast growth of the Higgs
boson expectation value toward the true vacuum drives the growth of gauge
fields, which start to contribute significantly after the mean value of
the Higgs field has first reached the vacuum expectation value (compare with
Fig.~\ref{means.conf}).  Since the total energy is conserved during the
evolution, the energy stored in the scalar fields decreases due to the
nonzero gauge energy fraction. This affects mostly the kinetic and
gradient energies of the scalar fields, while the potential energy remains
essentially unaltered.  This is also illustrated in
Fig.~\ref{ener.conf}, where we show the comparison of the average values
of all energy fractions for various values of the gauge coupling. From
the figure one notices a peculiar oscillatory dependence of the pure
gauge energy fraction with respect to $\gw$.  We have actually analyzed
this behavior in finer detail by running a few configurations for all
values of $\gw$ from $0.01$ to $0.2$ in intervals of $0.01$. The results
will be commented upon later in Section IV-C, in relation to
Chern-Simons production.

The previously discussed quantities provide an excellent testing ground
for checking the independence of our results on the various cutoffs
introduced in our procedure. As we discussed previously, the value of
these cutoffs has to be chosen judiciously to guarantee the
physical-character of the results. Figure \ref{means} tests the
dependence of the mean values of $\langle|\overline{\phi}|/v\rangle$,
$\langle \overline{\chi}/\chi_c \rangle$, as well as electric and magnetic
energies, on both $m a$ and $\pmin$ for model A1 (Table \ref{models}).
The mean values result from averaging over space as well as over initial
conditions.  The dispersion over the different set of initial conditions
gives rise to errors which are very small on the scale of the Figure.
Up to $ma=1.3$ we see no dependence at all of the evolution of the mean
values on the ultraviolet cutoff. There is a slight dependence on the
physical volume for $\pmin=0.2 m$, more significant for electric and
magnetic energies than for $\langle|\overline{\phi}|/v\rangle$, $\langle
\overline{\chi}/\chi_c \rangle$, but the results very rapidly converge
as we increase the volume.

As in paper I we have also checked that our results do not depend
strongly on the choice of the initial time $t_i$ at which the classical
evolution starts. 
All the results presented in this paper
correspond to the choice $mt_i=5$. In Fig. \ref{means_tidep} we show a
comparison of the evolution of spatial averages and the electric and
magnetic gauge energy fractions for $mt_i=5$ and $mt_i=6.5$ (for model
A1).  Most of the difference can be compensated by an additional shift
of size $-0.3$ in time. It is clear that a change in $t_i$ is well
accounted for by the change of initial conditions (very different in
magnitude), chosen according to the quantum linear evolution. This
proves the consistency of our analysis.

%%%%%%%%%%%%%%%  subsection spectra
\subsection{Energy behavior and spectra}

As discussed in Section \ref{classical}, the validity of the classical
approximation rests on the fact that infrared modes dominate the
dynamics of symmetry breaking and generation of Chern-Simons number.
The reliability of this approximation during the relevant time scales
has been challenged in Ref.~\cite{moore}. Moore has argued that,
once gauge fields are coupled to the Higgs-inflaton system, it is the
energy transfer between ultraviolet and infrared modes that drives the
dynamics, long before the physically relevant processes have taken
place. This would make it impossible to obtain a cutoff-independent
description of the evolution of the system and would invalidate the use
of the classical approximation. We will show in what follows that our
results seem to be free of this criticism. Essential for this is our choice
of initial conditions for the subsequent classical evolution. A similar
conclusion was found in Ref.~\cite{SST}, which use analogous initial
conditions.

The crucial test to address Moore's criticism is to check the
independence of the results of the UV cutoff at all the relevant time
scales. We have already tested this cutoff independence for average
mean values and energies in the previous Section. Here we will extend
the analysis to the Fourier spectra of the (gauge invariant) energy
densities. This Fourier decomposition exposes in a more direct way the
range of relevant momenta.  Figure \ref{spectra} shows, as an
illustration, the time evolution of the spectra of the electric
component of the gauge field energy density and of the Higgs-inflaton
contribution to the potential energy density. A similar behavior is
observed for the remaining terms (magnetic, kinetic, and gradient
parts for the Higgs and inflaton fields) of the Hamiltonian. Results
are presented for three different values of the lattice spacing:
$ma=1.30$, $0.87$, and $0.65$.  The results scale to an impressive
degree, and only in the last stages of evolution do deviations start
to be observed in the high momentum part of the distribution.  In Fig.
\ref{spectra_t45} we display the same Fourier spectra for $mt=45$.  An
enhancement of UV modes at the edge of the distribution is clearly
observed. As time evolves this altered distribution affects
lower-momentum modes as well, and tends to flatten the spectrum,
creating an artificial thermalization of the system. This effect is
already noticeable in the potential energy spectrum for $ma=2.60$, and
only slightly in that of $ma=1.30$.  Fortunately, as shown in Fig.
\ref{spectra}, in our physically relevant time interval and for the
actual values of the UV cutoff used in our analysis, we are free from
this problem.  As we will see in the next Section, the results on
Chern-Simons number generation support this conclusion.

%%%%%%%%%%%%%%%  subsection chern-simons

\subsection{Chern-Simons production} \label{cherns}

In this Section we will display  our results concerning Chern-Simons number
production.  We measure the variation in Chern-Simons
number between the initial time and a later time $t$:
\be
\label{DNCS}
\DNCS = \frac{\gw^2}{16 \pi^2} \int_{t_i}^{t} dt \int d^3\vec{x}\;  
{\rm Tr}(F_{\mu \nu}\tilde F^{\mu \nu}) \equiv \frac{1}{16 \pi^2}
\int_{t_i}^{t} dt \int d^3\vec{x}\; Q(\vec{x},t) \quad , 
\ee
where the dual tensor is   $\tilde F^{\mu\nu} = \half
\epsilon^{\mu\nu\rho\sigma} F_{\rho\sigma}$.
Since our dynamics is CP conserving, the average value of $\DNCS$ over
many configurations is zero.  We can expect that the mechanism of
Chern-Simons number production is local in space-time and characterized
by a certain physical length scale.  In such a situation it is more
interesting to express the result in terms of the sphaleron rate
$\Gamma$, given by
\be
\Gamma(t)= \frac{m^{-4}}{{\cal V}}\; \frac{d \langle 
\Delta N^{\ 2}_{\mbox{\tiny CS}}(t) \rangle }{dt}
\ee
which is independent of the physical volume ${\cal V}$ of the system. 
The symbol $ \langle \ldots \rangle $ denotes the expectation value with
respect to configurations obtained with different initial conditions
(i.e.  different random realizations). The factor $m^4$ is introduced to
make $\Gamma$ dimensionless. In our nonstationary situation, in which
the sphaleron rate might be time dependent, it is better to display the
quantity:
\be
I(mt)= m\int_{t_i}^{t} dt \, \Gamma(t) 
\ee

\begin{figure}[htb]
\vspace*{8.5cm}
\includegraphics{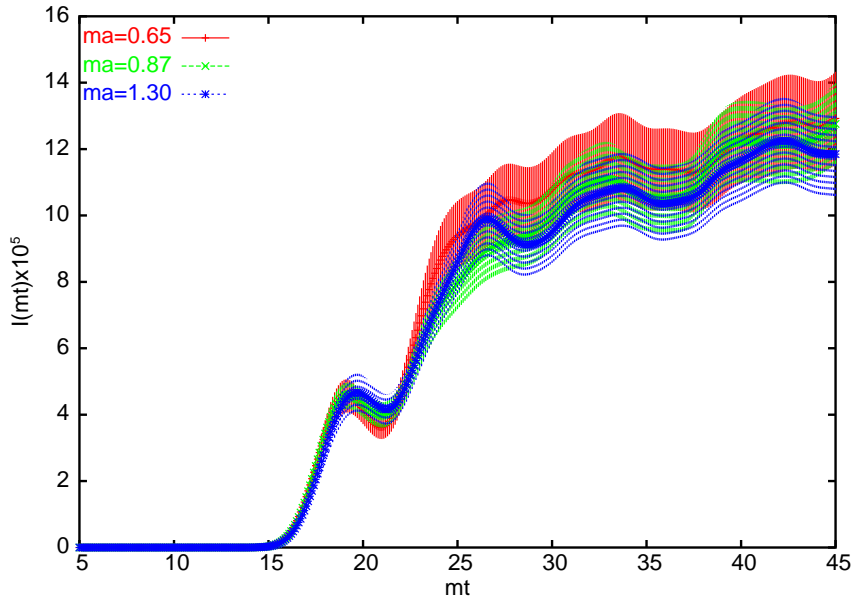}
\caption{ The time evolution of $I(mt)$ for model A1, $\pmin=0.15 m$ and
three different values of the lattice spacing.}
\label{ChernsimA}
\end{figure}

Our lattice implementation of the topological charge density $Q(\vec{x},t)=4
{\rm Tr}(\vec{E}\cdot\vec{B})$ is based on the discretized definitions of the 
electric and magnetic fields in terms of temporal and spatial plaquettes,
respectively. We have used $1\times1$,  $1\times2$ (for the electric field) 
and  $2\times2$ (for the magnetic field) plaquettes combined  in such a way 
as to subtract the leading ${\cal O}(a^2)$ correction for smooth fields. 
As we will see, our gauge fields vary over scales of the order of $1/\mw$,
which extend over several lattice spacings. Our {\em improved} discretization
of the topological charge density guarantees that we do not introduce
additional ${\cal O}(a^2)$ corrections in our Chern-Simons measurement. 
The situation contrasts with the one encountered in measurements of the
sphaleron rate at thermal equilibrium, where modes of the 
order of the cutoff are excited. We address the reader to the extensive
literature on the subject~\cite{sphalrate}.

As we will show later in this Section, in contrast to other quantities
described in the previous Subsections, the Chern-Simons number
fluctuates considerably from configuration to configuration. It behaves
like a random variable with a fairly large dispersion value. Thus, to
give a sufficiently precise value of $I(mt)$, one has to generate and
average over many configurations. The required precision is partly determined
by the possibility of testing  the cutoff independence
of the results.  Concerning the latter issue it was established in
previous Sections that there is a window of values of $\pmin/m$ and $ma$
where the dynamics of the system is insensitive to the actual values of
these cutoffs.  However, the Chern-Simons number might be more
sensitive to a scale on which cutoff effects are more apparent.

Although the Chern-Simons production mechanism is expected to be quite
different from the typical sphaleron tunneling processes in the
vacuum, it is still natural to expect that the relevant scales
involved are of the order of the two physical scales given by the
Higgs and W boson masses ($\mh=\sqrt{2}m$, $\mw=m \gw/(2
\sqrt{\lambda})$).  Thus, insensitivity to the ultraviolet cutoff
demands that $\mh a$ and $\mw a$ be small. The first requirement is
hard to meet with our present limitations of computer memory and
power. To reach values of $\mh \sim \mw$, while keeping the
ultraviolet and infrared effects under control, would require $\mh a
\sim 0.4$ and $\pmin \sim 0.15$, which implies large lattices, $N_s
\gsim 148$. With our present capabilities, we can reach values of $\mh
a$ that are only slightly below unity. Reducing this value compromises
the volume independence of the dynamics, since the physical size of
the bubbles become comparable to our physical box size.  Fortunately,
we know that in the low-temperature tunneling process between
different vacua, sphalerons exist in the limit $\mh \rightarrow
\infty$.  Thus, in this case the only relevant cutoff-dependent
parameter is $\mw a$, which can be made small by tuning the gauge
coupling $\gw$ and $\lambda$.  This situation was indeed confirmed by
previous lattice studies of sphalerons~\cite{margapierre}. Our choice
of ultraviolet and infrared cutoff values will be made within the
region that was found adequate in that study: the range given by $\mw
L \ge 2.5$, $\mw a\le 0.40$ for $\mh = \infty$ to $\mw L \ge 3.8$,
$\mw a\le 0.60$ for $\mh=\mw$, where $L$ is the linear extent of the
lattice in physical units.  As we will see, our results confirm that
this is also the case in our situation.

In conclusion, in order to study the physical Chern-Simons production in
our dynamical process, we have considered several models having
different degrees of sensitivity to cutoff effects. The most favorable
case, model A1, on which most of our study was focused, has
$\gw=0.05$ and $\lambda=0.00675$, giving $\mh/\mw=4.65$. The purpose of
studying other models is two-fold.  On one hand we can test the
sensitivity of the production mechanism to the ratio of scales
$\mh/\mw$, as was done recently for the Abelian case~\cite{ST}. On the
other hand, our study is meant to be a pilot one, which will allow the
determination of the computer resource requirements for the study of
other specific models.  For the different models (see
Table~\ref{models}), volumes and lattice spacings we have generated
typically $\sim200$ configurations on which our results are based.  The
full list of models, cutoff values, and the exact number of
configurations on which our results are based are collected in
Table~\ref{chernsim}. As explained before, the linear size of our box,
in $1/m$ units, is determined by the value of $\pmin$, by the expression
$mL=2\pi /(\pmin/m)$. On the other hand, the value of the lattice
spacing in physical $1/m$ units is determined by the number of lattice
points in each direction $N_s$, by the formula $ma=mL/N_s$. This allows
us to test ultraviolet and infrared effects separately by changing these
two quantities ($\pmin$ and $N_s$) independently.

%Table~\ref{chernsim}.
\begin{table}
\begin{tabular}{||c|c|c|c|c|c|c|c|c|c||}\hline\hline
Model & $N_s$ & $\pmin/m$ & Confs.  & $I(20)\times 10^5$ &
$I(25)\times 10^5$
& $\tilde{I}(25)\times 10^5$  &  
$I(35)\times 10^5$  &
$I(45) \times 10^5$ &$\tilde{I}(45) \times 10^5$\\ \hline \hline
A1  &  64  &   0.12   &   134  &  3.24(0.36)  &    7.94(0.87)  &  8.22(0.88)   &  12.35(1.25)  &  13.77(1.49)  &  12.45(2.00)  \\ \hline  
A1  &  48  &   0.12   &   191  &  3.77(0.35)  &    7.59(0.72)  &  7.72(0.80)   &  10.52(1.02)  &  12.12(1.08)  &  11.98(1.06)  \\ \hline
A1  &  64  &   0.15   &   140  &  4.14(0.42)  &    9.39(1.05)  &  9.20(0.99)   &  11.41(1.25)  &  12.92(1.40)  &  13.47(1.45)  \\ \hline
A1  &  48  &   0.15   &   191  &  4.35(0.43)  &    7.97(0.80)  &  7.46(0.48)   &  10.55(1.02)  &  12.76(1.22)  &  11.22(1.25)  \\ \hline
A1  &  32  &   0.15   &   167  &  4.61(0.56)  &    8.60(0.94)  &  8.54(0.72)   &  10.50(1.13)  &  11.85(1.26)  &  11.37(0.95)  \\ \hline
A1  &  48  &   0.2   &   152  &  5.91(0.49)  &    6.36(0.64)  &  6.82(0.37)   &  8.92(0.91)  &  9.32(0.99)  &  9.47(1.21)  \\ \hline \hline   
B   &  48  &   0.12   &   188  &  6.27(0.74)  &    12.49(1.41)  &  10.81(1.03)   &  13.86(1.54)  &  17.19(1.99)  &  14.70(1.47)  \\ \hline
B   &  64  &   0.15   &   182  &  8.10(0.92)  &    13.86(1.22)  &  15.16(2.26)   &  17.16(1.45)  &  19.89(1.74)  &  19.60(3.26)  \\ \hline
B   &  48  &   0.15   &   200  &  7.46(0.80)  &    12.77(1.33)  &  11.65(1.23)   &  14.74(1.39)  &  17.79(1.68)  &  15.11(1.67)  \\ \hline 
B   &  32  &   0.15   &   170  &  8.26(0.92)  &    10.83(1.19)  &  7.93(0.98)   &  13.29(1.46)  &  14.89(1.65)  &  10.84(1.31)  \\ \hline \hline
A2  &  48  &   0.15   &   210  &  5.41(0.60)  &    17.63(1.62)  &  17.39(1.22)   &  25.16(2.54)  &  33.87(3.61)  &  24.79(2.20)  \\ \hline 
A3  &  48  &   0.15   &   128  &  3.52(0.41)  &    19.46(2.67)  &  14.72(1.50)   &  28.71(4.25)  &  38.00(5.62)  &  26.22(2.68)  \\ \hline 
A3  &  32  &   0.15   &   181  &  3.58(0.42)  &    14.68(1.64)  &  12.85(1.35)   &  20.80(2.08)  &  24.58(2.42)  &  22.92(2.02)  \\ \hline 
A4  &  48  &   0.15   &   157  &  0.54(0.06)  &    5.06(0.52)  &  4.72(0.41)   &  10.03(1.07)  &  16.10(1.74)  &  14.69(1.49)  \\ \hline 
A4  &  32  &   0.15   &   181  &  0.70(0.06)  &    6.49(0.72)  &  5.56(0.43)   &  10.12(1.00)  &  14.17(1.44)  &  13.71(1.44)  \\ \hline 
\end{tabular}
\caption{Table of results specifying the model, the lattice size $N_s$,
$\pmin/m$, number of configurations, and values of $I(mt)$ for various 
times ($mt=20,25,35,45$). The symbol  $\tilde{I}(mt)$ stands for  $I(mt)$ as
determined by fitting the histogram of $\DNCS$ to a normalized Gaussian. }
\label{chernsim}
\end{table}

\begin{figure}[htb]
\vspace*{7.7cm}
\includegraphics{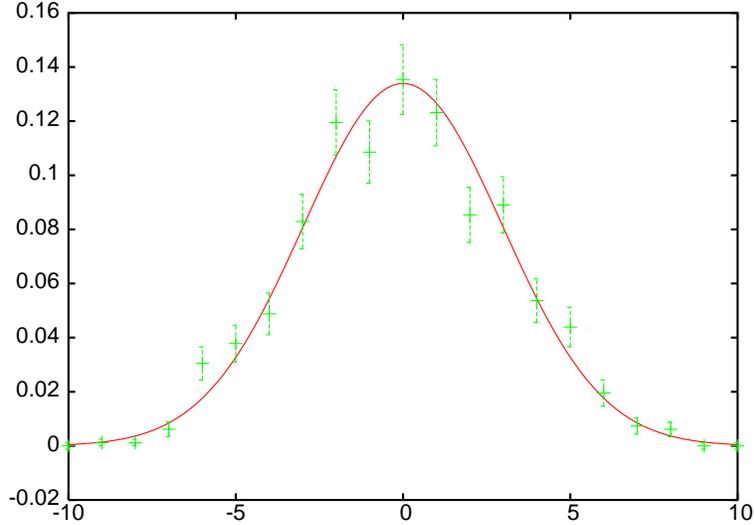}
\caption{ Combined histogram of values of $\DNCS$ for model A1, 
at $mt=45$, $\pmin=0.15m, 0.12m$ and
$N_s=32,48,64$. The data of $\pmin=0.12m$ are rescaled by the square root of
the volume.}
\label{histogram}
\end{figure}

\begin{figure}[htb]
\vspace*{7cm}
\includegraphics{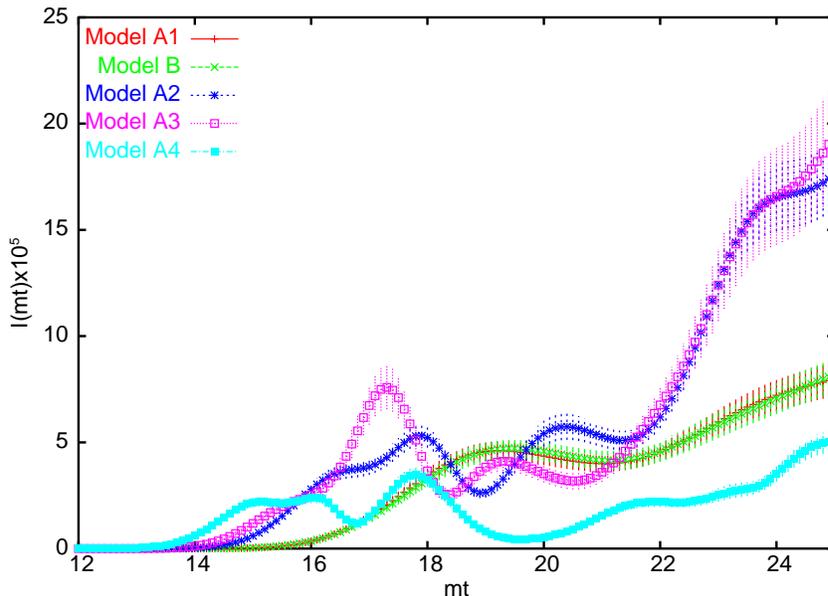}
\caption{ The time evolution of $I(mt)$ with  $\pmin=0.15 m$, $N_s=48$  for
all our models.}
\label{ChernsimB}
\end{figure}

We stress that volume and lattice-spacing dependences should affect
the data very differently. The distribution of Chern-Simons number
should not depend on the value of $N_s$ if ultraviolet cutoff effects
are small.  This is very well satisfied by our data on $\langle \Delta
N^{\ 2}_{\mbox{\tiny CS}}(t) \rangle$ as seen in Fig.~\ref{ChernsimA},
where we have displayed the time evolution of this quantity for model
A1, $\pmin=0.15m$ and various values of the ultraviolet cutoff
$N_s=32$ ($ma=1.31$), $N_s=48$ ($ma=0.87$), and $N_s=64$ ($ma=0.655$).
In this figure we observe that $I(mt)$ starts growing from zero, its
initial value, at a time $mt\approx 15$.  This growth is very steep
and presents oscillations up to a value of $mt\approx 25$. Then the
growth continues with a smaller slope. The insensitivity of our
results to the value of the ultraviolet cutoff is very striking given
the relatively large value of $ma$. It provides evidence that the
relevant cutoff scale is $\mw a$ (=$0.37$, $0.25$, $0.185$).

If the mechanism of Chern-Simons generation is {\em local in space}, one
expects that a change in the volume should modify the dispersion $\DNCS$
proportionally.  Therefore, as mentioned previously, the quantities
$\Gamma(t)$ and $I(mt)$ should be independent of the spatial volume.

Our results for $I(t)$ at various times ($mt=20, 25, 35$ and $45$) are
collected in Table~\ref{chernsim}.  The errors quoted within
parentheses are computed in the standard fashion from the dispersion
within the various configurations.  For a typical number of
configurations of approximately 150, they are around 10$\%$. For model A1
the results for various volumes are displayed, showing a large degree of
independence of $I(t)$ of $\pmin$. There is nevertheless some
significant dependence for the smallest volume ($\pmin=0.2$). This
dependence was also observed in mean values and energies. For all the
quantities the volume dependence decreases considerably for larger
volumes ($\pmin=0.15, 0.12$), being of the order of the quoted errors.

Furthermore, the cutoff independence of the data is an important and
nontrivial point.  Choosing values of the model parameters outside the
relatively small optimal range gives results that behave quite
differently (e.g., rising with $N_s$).  The fluctuating nature of
$I(t)$ forces one to accumulate a certain amount of data before the
cutoff dependence shows up clearly. This sends a word of warning to
other researchers studying similar questions.

In addition to studying the value and time dependence of $I(t)$, it is
interesting to investigate the distribution of values of the
Chern-Simons number $\DNCS$ for various times. In the absence of a CP
violating term this distribution should be centered at $0$ and have a
dispersion determined by the value of $I(t)$.  Assuming, as we did
before, that the Chern-Simons number is produced in local structures
with arbitrary signs, one expects the distribution to be of the Poisson
type. This, for large number of structures, is close to a Gaussian
distribution centered at zero. Indeed, our results agree with these
expectations. Histogramming, normalizing and fitting the data to a
Gaussian allows one to obtain an estimate of $I(t)$, the only parameter
of the fit.  The values for $t=25$ and $t=45$ are displayed in
Table~\ref{chernsim} (using the symbol $\tilde{I}(t)$.  The resulting
$\chi^2$ per degree of freedom was most of the time of the order of
1 and never larger than 2. This confirms that the distribution of
the Chern-Simons number over the different configurations is compatible
with a Gaussian. In Fig.~\ref{histogram} we show a combined histogram of
our data for model A1, $mt = 45$, $\pmin = 0.12, 0.15$, and all values of
$N_s$. The normalized Gaussian that best fits the data is also
displayed. The $\chi^2$ of the fit is $1.2$ per degree of freedom, and
the value of $I(45)$ resulting from the fit is $12.07(\pm0.64)
\times10^{-5}$, which is perfectly compatible with the data in
Table~\ref{chernsim}.

Finally, we will comment on the results for other values of the
parameters of the model. It is obvious from Table~\ref{chernsim} that
the results are of the same order of magnitude, but might differ by
factors of 2 or larger.  In order to make a direct comparison one
should take into account that the time evolution of $I(mt)$ is quite
different for all the models.  This is shown in Fig.~\ref{ChernsimB}
where we display $I(mt)$ for $N_s=48$ and $\pmin=0.15m$ for all the
models and a relevant range of times. The data corresponding to model
B are shifted in time by $3.3m$ and scaled up by $20\%$. With this
modification they match perfectly with those of model A1. These two
models have a common value of $\mh/\mw$.  We find this agreement
remarkable, since model B is in a region of parameters that is
expected to be more sensitive to ultraviolet and infrared cutoffs. On
the other hand all A-models have the same value of $\lambda$ and
different values of $\gw$, and hence different ratio of masses.  It is
clear that the evolution of $I(mt)$ has a fairly sensitive dependence
on $\mh/\mw$.  A similar strong dependence of the Chern-Simons number
on $\mh/\mw$ has been reported recently for the (1+1)-dimensional Abelian
model~\cite{ST}.  We found a certain positive correlation between the
values of the gauge-energy fraction and that of $I(mt)$. Since the
former quantity has fewer fluctuations and requires fewer statistics,
we studied a few configurations for all values of $\gw$ ranging from
$0.01$ up to $0.2$ in intervals of $0.01$ and monitored the fraction
of gauge energy at $mt=45$. The result shows an oscillation, growing
from $\gw=0.01$ to a maximum at $\gw=0.0825$ (model A2), and then
decreasing up to $\gw=0.147$, where it begins to grow again. We
therefore expect that the maximum values of $I(mt)$ attainable in this
region of parameters are not far from those of model A2.

%\subsection{Space-time structures}

%%%%%%%%%%%%%%%  subsection structures

\subsection{Space-time structures}

In this Subsection we will investigate the local structures associated
with Chern-Simons number production. Since in the first stages of
evolution and for our choice of inflaton coupling ($g^2=2\lambda$),
the Higgs and the inflaton fields evolve collinearly, we will restrict
our analysis to Higgs and gauge field structures.

Initially the space-time structures of the scalar fields do not differ
significantly from those of paper I, in which gauge fields are
decoupled.  Snapshots of the evolution of the Higgs field from the
growth of spatial lumps to the generation of bubbles were presented
there, where we also gave a partial analytical description of the
process.

\begin{figure}[htb]
\vspace*{9.0cm}
\includegraphics{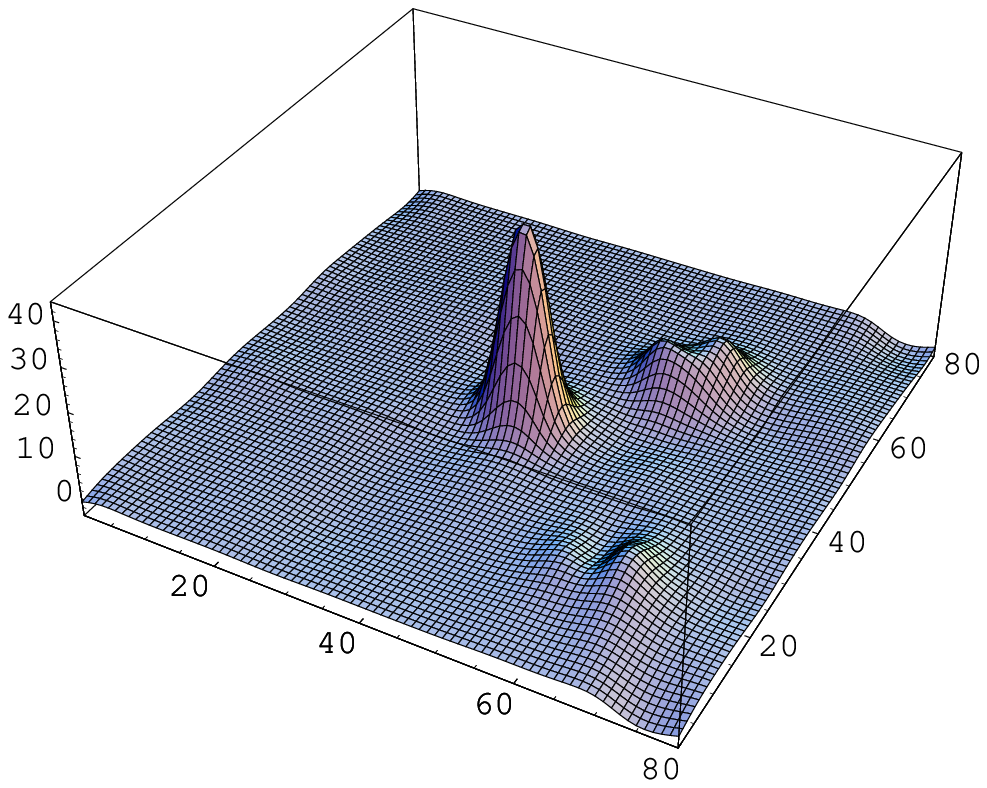}
\includegraphics{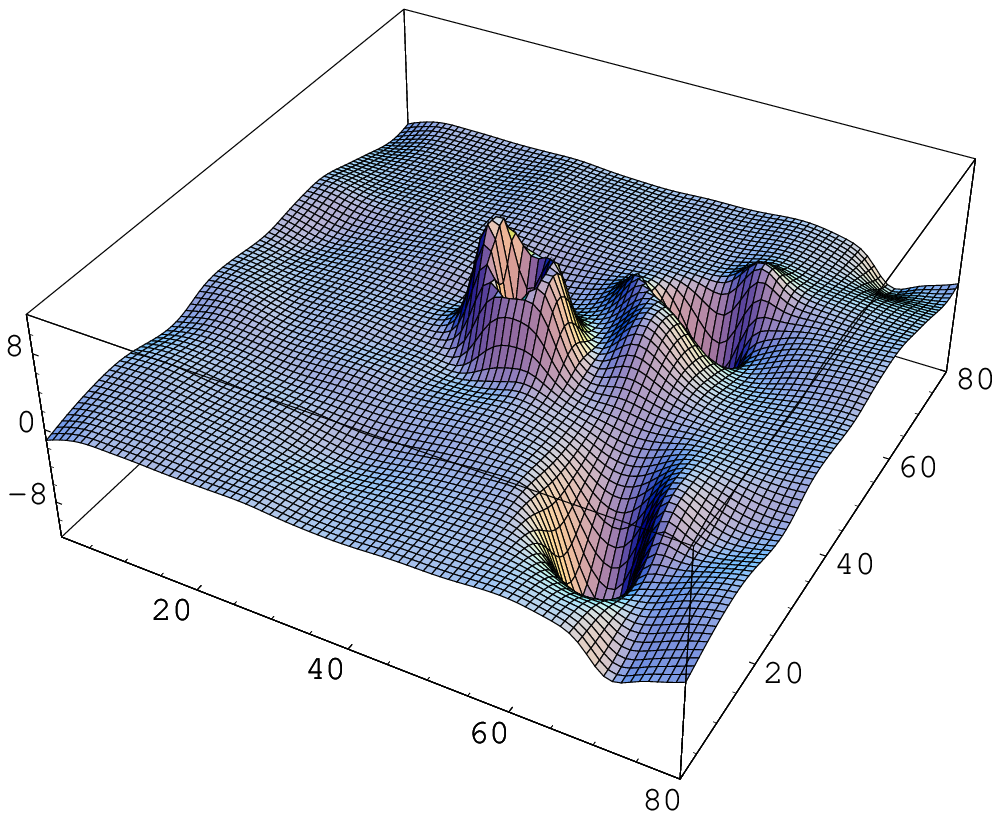}
\caption{Snapshots of $Q(\vec x,t)$ at, from left to right,
$mt= 18$  and $mt=19$ (model A1, $N_s=80$ and $\pmin=0.15m$). Contours
> are shaded such that red (dark in black and white display) represents
> the false vacuum, while the true vacuum is yellow (light in black 
> and white display).}
\label{eb_80}
\end{figure}

\begin{figure}[htb]
\vspace*{8.5cm}
\includegraphics{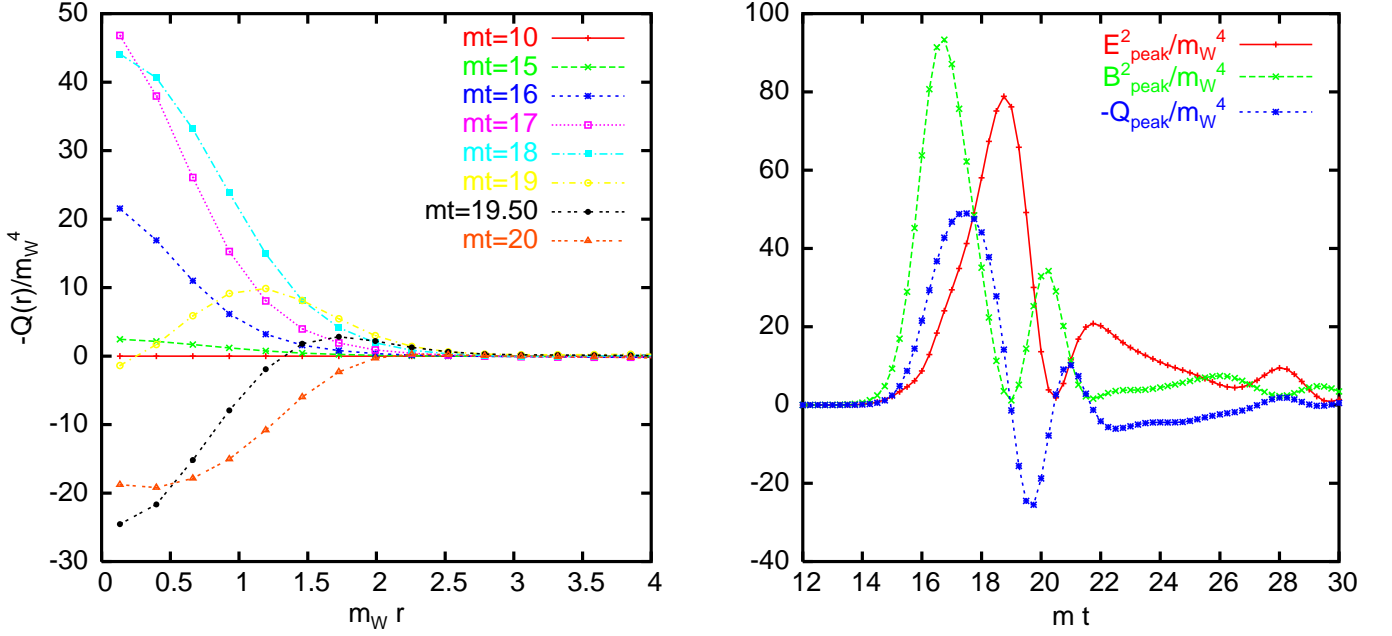}
\caption{Left: The spatial profile of the charge density peak for
various times. Right: The value of the electric, magnetic and charge density
at the center of the peak as a function of time.}
\label{perfiles}
\end{figure}

Our main interest now concerns the structure of the topological charge
density $Q(\vec x, t)$, which integrated over space and time gives
rise to $16\pi^2\,\DNCS$, as described in the previous Section. For
the case of model A1, $I(mt)$ starts to become sizable at $mt \gsim
15$. The local origin of this growth is well exemplified by the
snapshots of $Q(\vec x, t)$ on two dimensional slices presented in
Fig.~\ref{eb_80}. The slices were chosen to coincide in one coordinate
with a local maximum of the topological charge density. Lumps such as
the ones observed in the figure become manifest for $mt \gsim 15$.
During the following stages of evolution, their position remains
unchanged but the value at the peak presents an oscillatory pattern,
as was the case for the Higgs boson lumps, even flipping the sign of
the charge density at the center of the lump.  This is illustrated in
Fig.~\ref{perfiles} (left) where the charge density profile is shown
at different times for the same configuration displayed in
Fig.~\ref{eb_80}.  The origin of this oscillation seems to be
associated to a similar oscillatory behavior of the electric and
magnetic fields.  This can be deduced from their behavior at the
center of the charge density peak as a function of time which is
displayed in Fig.~\ref{perfiles} (right).  Electric and magnetic
fields oscillate out of phase, and changes in the sign of the charge
density seem to take place when the modulus of either the magnetic or
the electric field vanishes.  The flip of sign of the charge density
could be a possible explanation for the oscillation in $I(mt)$
observed around $mt=20$ in Fig.~\ref{ChernsimA}.

\begin{figure}[htb]
\vspace*{8.5cm}
\includegraphics{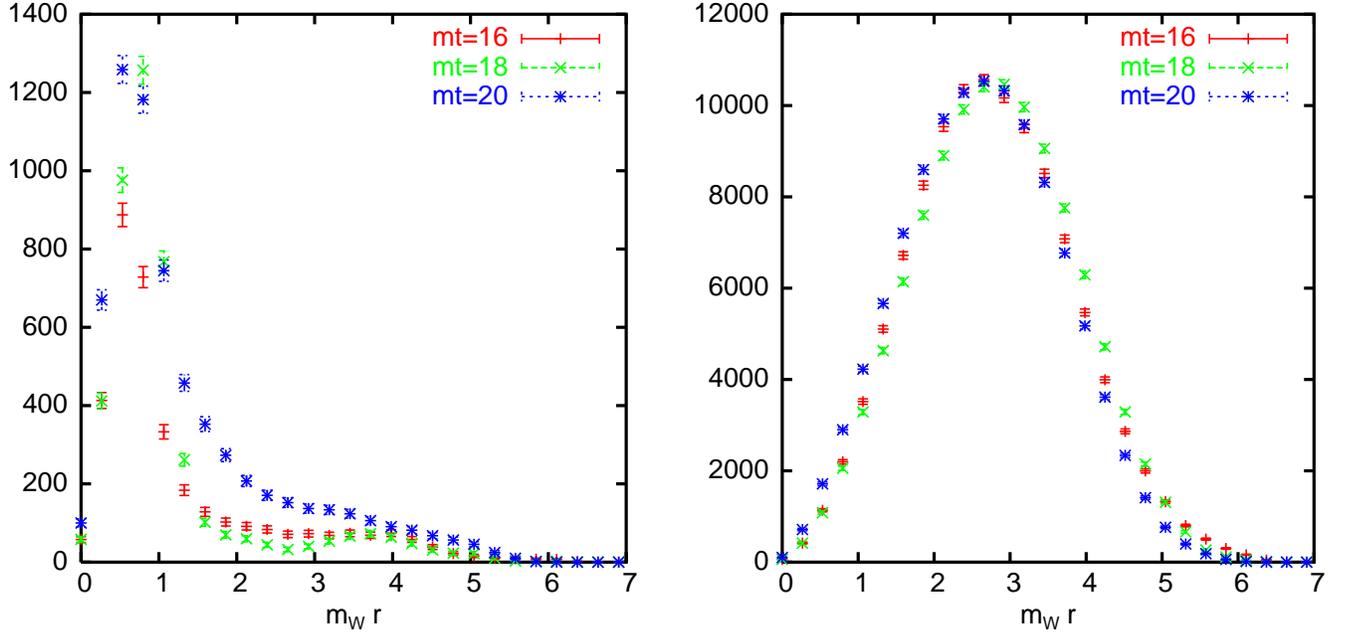}
\caption{Histograms of minimal distance to a peak of the charge density
from: points in $\Lambda_0$ (left); all lattice points (right).
Data for model A1, $N_s=48$ and $\pmin=0.15m$.}
\label{histo_dist}
\end{figure}

To provide a more systematic statistical study of the local structure of
the topological charge density, we have performed a detailed study of a
few configurations of model A1 with $N_s=48$ 
and $\pmin=0.15$.  The data show a remarkable spatial
concentration of the density. For example, selecting the lattice points
($\Lambda_0$) for which $|Q(\vec x, t)| \ge 2.33 \ \mw^4$, we find that
they occupy only a few percent of the lattice volume ($3\%$, $4\%$ and
$6\%$ for $mt=16$, $18$, and $20$, respectively), while they account for a
large fraction of the total charge squared ($77\%$, $92\%$, and $
83\%$).

Another way to investigate clustering is by identifying local maxima (or
``peaks'') of the distribution $|Q(\vec x, t)|$, with height above a
certain threshold ($|Q(\vec x, t)| \ge 7 \mw^4$).  The number of these
peaks starts to be nonzero for $mt\gsim 15$, about the time when we
start to observe a fast growth in the dispersion of the Chern-Simons
number. Beyond this time, and up to $mt\sim 25$, 
there are typically a few ($\lsim 10-20$ ) such maxima.  Most of the points in
$\Lambda_0$ are clustered around these peaks, as shown by the histogram
of minimum distances from points in $\Lambda_0$ to the peaks
Fig.~\ref{histo_dist} (right). The shape has to be compared with the histogram
of minimum distance from any lattice point to one of the peaks, which
gives a measure of phase-space, see Fig.~\ref{histo_dist} (left).  From the
comparison one also concludes that the majority of points which are
sufficiently close to one of the peaks belong to $\Lambda_0$. Thus,
peaks identify structures of the size of a few lattice spacings,
compatible with $\lsim 1/\mw$ in physical units.

\begin{figure}[htb]
\vspace*{15.3cm}
\includegraphics{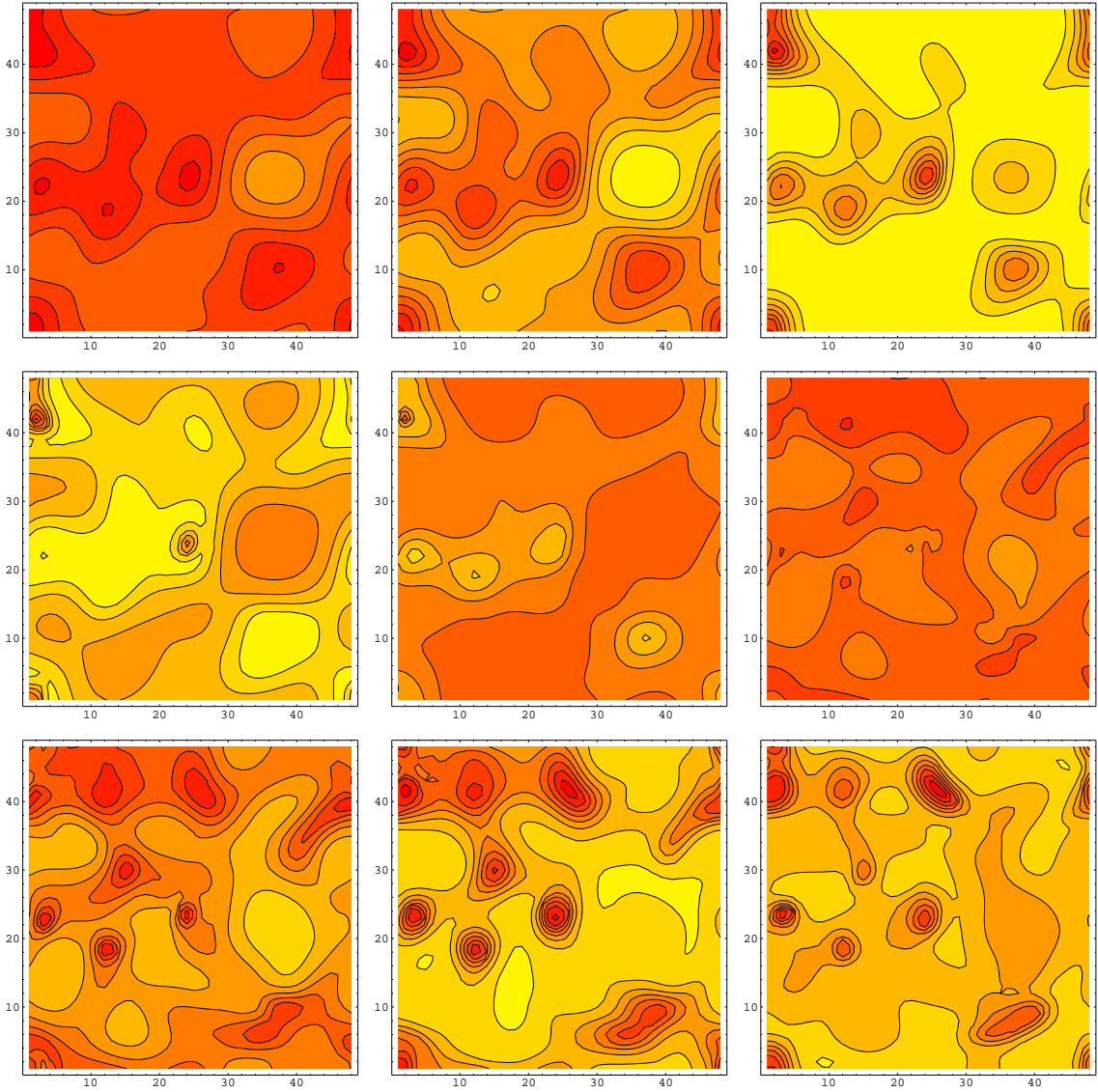}\caption{Contour plots of $|\phi(x)|^2$ at Top:  $mt=12$, $mt=13$ and
$mt=14$, Middle:  $mt=15$ $mt= 16$ and $mt=17$
and Bottom: $mt=18$ $mt= 19$ and $mt=20$.
model A1, $N_s=48$ and $\pmin=0.15m$.}
\label{contour}
\end{figure}

\begin{figure}[htb]
\vspace*{15.3cm}
\includegraphics{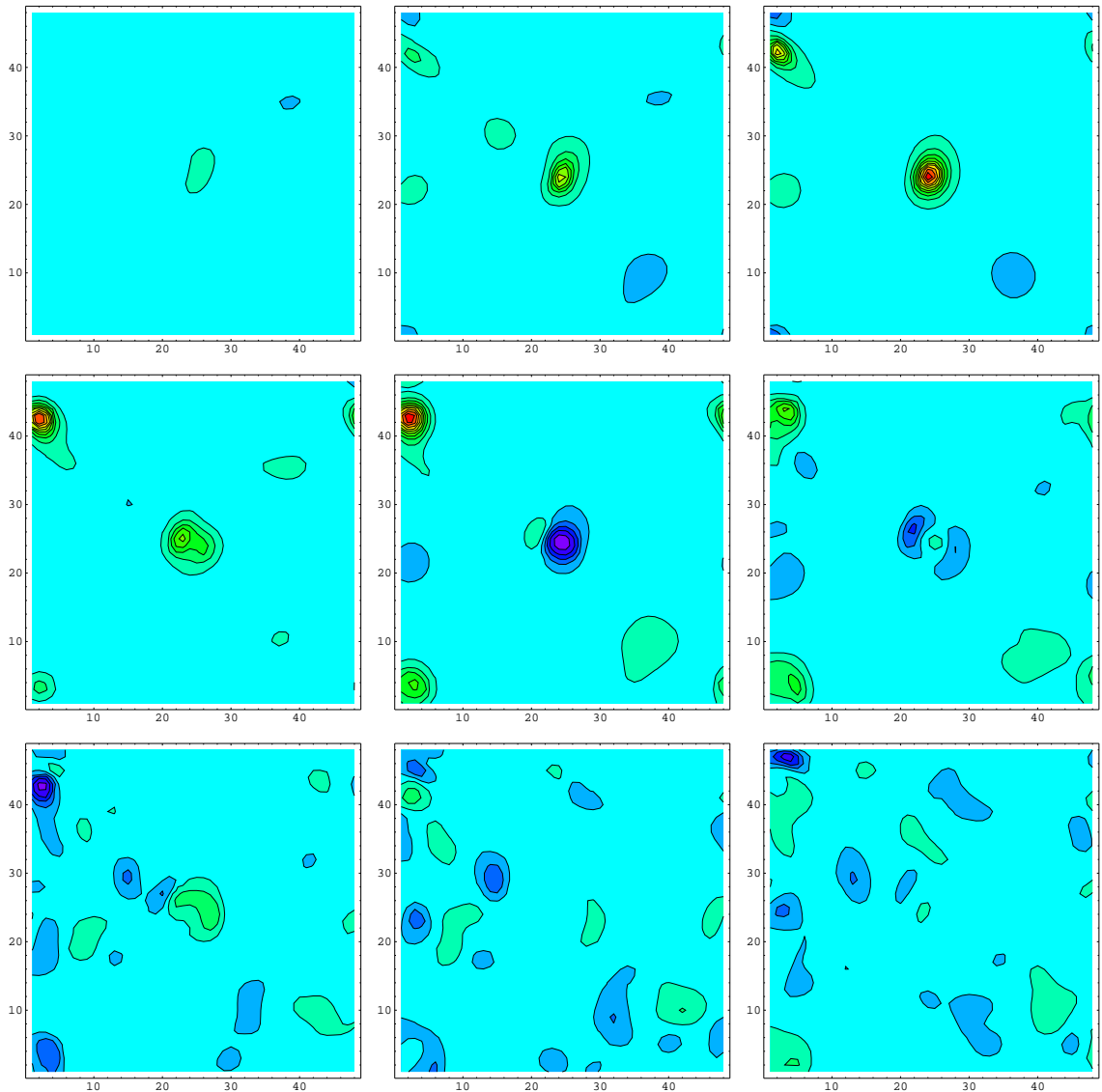}\caption{Contour plots of $Q(\vec x,t)$ at
Top:  $mt=15$, $mt=16$ and $mt=17$, Middle: $mt=18$, $mt=19$
and $mt=20$ and Bottom: $mt=21$, $mt=22$ and $mt=23$. model A1, $N_s=48$ and
$\pmin=0.15m$. Contours are shaded such that lumps in the topological charge
> density appear as red regions (dark in black and white display).}
\label{contour2}
\end{figure}

A similar conclusion can be drawn by computing the fraction of charge
density square contained inside spheres of radius $1/\mw$ around each
peak.  This amounts to $ 60 \%$, $82 \%$ and $ 64\%$ at $mt=16$, $18$,
and $20$ respectively, while these spheres occupy only $3.5\%$, $3\%$,
and $4.6\%$ of the total lattice volume.  Summarizing, we conclude
from the results of this analysis, that the topological charge density
is concentrated in local structures of typical size $1/\mw$, such as
those depicted in Fig.~\ref{eb_80}.  Similar structures are also
observed in the electric and magnetic energy densities.

In view of these results one might be tempted to identify the local
structures appearing in the first stages of evolution with
sphaleronlike configurations \cite{KM}.  Sphalerons are static
solutions of the SU(2)-Higgs classical equations of motion with
fractional Chern-Simons number.  They have indeed a localized, lumpy,
energy density with a typical size given by $1/\mw$. However, it is
essential for the sphalerons that lumps in the energy density are
correlated with zeros in the Higgs field, i.e.  $\vec B^2 (x)$ gets a
maximum at the location where $|\phi (x)|=0$.  To take a closer look
at the correlation between the structures of the Higgs field and of
the topological charge density in our configurations, we present in
Figs.~\ref{contour} and \ref{contour2} contour plots of $|\phi(x)|$
and $Q(x)$ on two-dimensional slices for various values of $mt$.

Let us first analyze the Higgs contours on Fig.~\ref{contour}.
Contours are shaded such that the false vacuum ($|\phi(x)|=0$) is
represented by red (dark in black and white display), while the true
vacuum ($|\phi(x)| \gsim 1$) is yellow (light in black and white
display).  At $mt=12$ we observe the formation of the first Higgs
lumps [medium grey (orange) in the figure], which grow from local
maxima in the initial random Gaussian distribution of the Higgs field.
These lumps grow, reach the minimum of the potential, and start to
oscillate around it; see $mt=13$, $14$, and $15$.  As observed in
Fig.~\ref{contour} there are a series of very localized regions
between lumps which remain in the false vacuum [dark (red)] for a
longer time and only much later start to oscillate around the minimum
of the potential, almost opposite in phase to the oscillations of the
Higgs homogeneous zero mode~\cite{footnote}.  Remarkably, it is
precisely at these points where localized lumps appear in the
topological charge density, appearing as red regions (dark in black
> and white display) in
Fig.~\ref{contour2}. Notice, however, that there is a relative time
shift $\Delta mt= 2-3$ between the presence of zeroes in the Higgs
field and that of maxima in the charge density.

In summary, we have shown that there is a clear association between 
the behavior of $I(mt)$ and the presence of local structures of 
size $\sim 1/\mw$ in gauge degrees of freedom, also correlated to 
certain structures in the Higgs field. A more detailed investigation of 
these issues is certainly interesting but demands more intensive studies 
which fall beyond the scope of this paper.

%%%%%%%%%%%%%%%%%%%%%%%%%%%%%%%%%%%%
%%%%%%    SECTION BARYOGENESIS
%%%%%%%%%%%%%%%%%%%%%%%%%%%%%%%%%%%%

\section{Baryogenesis}\label{Baryogenesis}

In this Section we will illustrate how the results of the 
previous ones can be related to baryon number generation.
In the Standard Model, baryon and lepton numbers are not conserved
because of the anomalous nonconservation of the associated left-handed
fermionic (leptonic or baryonic) currents, $j^\mu =\bar\psi_{_L}
\gamma^\mu\psi_{_L}$, through the chiral anomaly~\cite{tHooft}, 
\begin{equation}
\partial_\mu j_B^\mu = \partial_\mu j_L^\mu =
{3\over16\pi^2}\,Q(\vec{x},t) \, = \, 
3\partial^\mu K_\mu\,,
\end{equation}
where $Q(\vec{x},t)$ is the topological charge density studied in the
previous Section, and $K_\mu$ is the Chern-Simons current. In this way
we see how the change in baryon and lepton numbers are related to that
of the Chern-Simons number: $\Delta B = \Delta L = 3 \DNCS$.

At zero temperature variations in Chern-Simons number can proceed
through a tunneling process~\cite{tHooft} with negligible
probabilities $\Gamma \sim \exp(-4\pi/\alphaw) \sim 10^{-170}$. We
will refer to those transitions (quantum or classical) that change
Chern-Simons number as sphaleron transitions.

At nonzero temperatures thermal fluctuations can help the generation
of hot sphalerons, allowing over-the-barrier transitions between
different vacua that may be responsible for baryon
production~\cite{KRS}. Below the critical temperature they are still
exponentially suppressed, but with a temperature-dependent rate,
$\Gamma \propto \exp[-E_{\rm sph}(T)/T]$, where $E_{\rm sph}(T)$ is
the sphaleron energy. However, at high temperatures, the sphaleron
transitions are mainly sensitive to the long-wavelength modes in the
hot plasma. A simple argument then suggests that the rate of sphaleron
transitions per unit time per unit volume should be of the order of
the fourth power of the magnetic screening length in the
plasma~\cite{armcl,khsh}.

Away from equilibrium the transition rate is harder to estimate.
However, in Ref.~\cite{GGKS} it was proposed that the sphaleron
transition rate during rescattering after preheating, $\Gamma_{\rm
  sph}$, could be high, and could be estimated as that of a system in
local thermal equilibrium at some effective temperature $\teff$ for
the long-wave modes, $\Gamma_{\rm sph} \approx \alphaw^4 \teff^4$,
characteristic of the equipartition energy stored in those modes, $E
\sim k_{\rm B} \teff$.

The absence of a CP-violating interaction in our model implies that
the average value of $\DNCS$ must be zero and no net baryon number can
be created on the average. However, there is a scenario for
incorporating a CP-violating effect proposed in Ref.~\cite{GGKS},
which will allow us to connect the results presented in the previous
Sections with the issue of EW baryogenesis. This will be explained in
what follows.

One possible way to induce a CP-asymmetry within the effective field
theory approach, is to consider the inclusion of nonrenormalizable
operators that violate CP.  These could be low energy remnants of
additional degrees of freedom and interactions present at a higher
energy scale $M_{\rm new}$.  The lowest dimension-6 operator of this
sort is~\cite{misha}
\begin{equation}\label{cpnonc}
{\cal L}_{\rm CP}  = \delcp\,{\Phi^\dag\Phi \over M_{\rm new}^2}\,
{3\gw^2\over32\pi^2}\,F^a_{\mu\nu}\tilde F^{\mu\nu}_a \,,
\end{equation}
which is both P and CP violating. The dimensionless parameter $\delcp$
is then an effective measure of CP violation.  The factor $3\gw^2/32
\pi^2 \sim 4\times10^{-3}$ is introduced in order to simplify the
subsequent formulas.  Note that the operator (\ref{cpnonc}) does not
violate C, and therefore in the bosonic sector of the theory the
nonequilibrium evolution can produce only P- and CP-odd configurations.
The required C violation comes from the fermionic sector of the
theory, which has not been included in our work.  These C-violating
operators appear in gauge-fermion EW interactions that violate C and
P, but conserve CP.

Following Ref.~\cite{GGKS}, we will assume that  the operator
(\ref{cpnonc}) induces an effective chemical potential $\mueff$,
which introduces a bias, i.e. a slope, in the potential of EW vacua,
between baryons and antibaryons~\cite{GGKS}, 
\begin{equation}\label{mueff}
\mueff = {\delcp\over M_{\rm new}^2}\,
{d\over dt}\langle\Phi^\dag\Phi\rangle.
\end{equation}
Although the system is very far from thermal equilibrium, we will
assume that the collision integral equation governing the time
evolution of the baryon number density can be cast in the form of a
Boltzmann-like equation, where only the long-wavelength modes
contribute~\cite{misha,GGKS}
\begin{equation}\label{Boltzmann}
{d\,\nb\over dt} = \mueff\,{\Gamma_{\rm sph}\over\teff} - 
\Gamma_{_{\rm B}}\,\nb,
\end{equation}
where $\mu_{\rm eff}$ is the time-dependent effective chemical
potential~(\ref{mueff}), and $\Gamma_{_{\rm B}}={39\over2}\Gamma_{\rm
  sph}/\teff^3\ll m$ is the term responsible for baryon
washout~\cite{khsh}.  As can be deduced from Fig.~\ref{ChernsimA}, the
sphaleron rate $\Gamma_{\rm sph}$ grows very quickly initially but
then decreases to a much lower value at later times.  On the other
hand, the effective temperature $\teff$ decreases because of
rescattering, while the energy stored in the long-wave modes is
transferred to the high-momentum modes~\cite{GGKS,FK,GBKLT}. The
washout rate $\Gamma_{_{\rm B}}$ is smaller, even at high
temperatures, than the other scales.  For instance, for $\teff=400$
GeV and $\Gamma_{\rm sph} \sim 2\times 10^{-5}\,m^4$ we obtain
$\Gamma_{_{\rm B}}=0.01$ GeV, which is small compared with the
tachyonic Higgs growth toward symmetry breaking ($m/10\sim30$ GeV).
It is also much smaller than the perturbative decay rates of the Higgs
boson into gauge fields, or those into light fermions. Therefore, we
will assume that the last term in Eq. (\ref{Boltzmann}) is negligible
during baryon production immediately after EW symmetry breaking, and
the final baryon asymmetry can thus be estimated by integrating
Eq.~(\ref{Boltzmann}):
\be\label{baryon}
\nb = \int dt\,\Gamma_{\rm sph}\,{\mueff\over\teff} \simeq
\Gamma_{\rm sph}\,{\delcp\over\teff}\,
{\langle\Phi^\dag\Phi\rangle\over M_{\rm new}^2}\,,
\ee
where all quantities are evaluated at the time of symmetry breaking
(more precisely at the maximum of the sphaleron rate).
This corresponds to a baryon asymmetry
\be\label{bau}
{\nb\over s} \simeq {45\,\delcp\over2\pi^2\,g_*}\,
{ v^2\over M_{\rm new}^2}\,
{\Gamma_{\rm sph}\over\teff\,\trh^3} \simeq
15\,\delcp\,\left({3\over2\pi^2\,g_*}\right)^{1/4}
\left({160\, {\rm GeV}\over\mh}\right)^{3/2}\,
{v^2\over M_{\rm new}^2}\,{\Gamma_{\rm sph}\over\teff\,v^3}\,,
\ee
where $g_* = 106.75$ is the number of effective degrees of freedom that
contribute to the entropy density, $s = (2\pi^2\,g_*/45)\,\trh^3$, at
the electroweak scale. Taking the scale of new physics to be at 
$M_{\rm new}\sim 3$ TeV, the Higgs boson mass $\mh\sim160$ GeV, the 
sphaleron rate $\Gamma_{\rm sph} \sim
2\times 10^{-5}\,m^4$ (see Table~\ref{chernsim}), and the effective
temperature $\teff \sim 400$ GeV, we find
\be\label{BAU}
{\nb\over s}  \simeq 1.6\times10^{-6}\,\delcp\,{v^2\over M_{\rm new}^2} 
\simeq 1.1\times10^{-8}\,\delcp\,,
\ee
consistent with observations for $\delcp\sim10^{-3}$, a perfectly
acceptable value for particle physics beyond the standard model.
Therefore, baryogenesis at EW preheating after hybrid inflation
\cite{GGKS} can be very efficient, in the presence of a new source
of CP violation. Varying the scale of the new physics $M_{\rm new}$,
one can satisfy the observational constraints by changing $\delcp$
accordingly.  

The previous calculation has to be understood as an order of magnitude
estimation of the baryon number production within the scenario of
Ref.~\cite{GGKS}, since there are clear uncertainties in both
$\Gamma_{\rm sph}$ (with realistic couplings) and $\teff$.
Furthermore, there are also theoretical caveats. On one hand, the
connection between Chern-Simons number and baryon number generation in
a nonperturbative setting as ours has been questioned in
Ref.~\cite{klinkhamer}. On the other hand, doubts have also been
expressed in Ref.~\cite{RSC} about the validity of the Boltzmann
equation~(\ref{baryon}) in a non-Markovian evolution of the
Chern-Simons number production. Modulo these warnings, we may conclude
that the observed baryon asymmetry does not require unnaturally large
values of $\delcp$.  Remarkably, a recent paper~\cite{ST2} that does
include the CP-violating operator (\ref{cpnonc}) in the simulations
finds a similar estimate of the baryon number generated within this
scenario, despite the use of different initial conditions (quench
approximation, etc.) and model parameters.

%%%%%%%%%%%%%%%%%%%%%%%%%%%%%%%%%%%%
%%%%%%    SECTION CONCLUSIONS
%%%%%%%%%%%%%%%%%%%%%%%%%%%%%%%%%%%%

\section{Conclusions}

In this paper we have studied the evolution with time of a quantum
field theory containing the SU(2) gauge-scalar sector of the standard
model coupled to a singlet inflaton. The initial state of the system
follows from the end of a hybrid inflation scenario, which is
dominated by a slowly decreasing homogeneous inflaton mode, coupled to
Higgs and gauge fields in the Minkowski vacuum. The tachyonic
instability of the model at the initial stages triggers the fast
growth of the infrared modes of the Higgs, which evolve toward
classical behavior.  This provides justification of our main
approximation: the assumption that the classical evolution of this
field leads to a classical behavior of the remaining modes of the
whole system, and that this occurs before the nonlinear effects,
including backreaction become important. The consistency and
justification of this idea were studied in paper I, in the absence of
gauge fields. Other authors have applied similar
ideas~\cite{PS,TK,GBKLT,CPR} in different contexts. In this paper, we
built upon this idea and incorporated gauge fields in the problem. We
have tested the self-consistency of the scheme by varying the initial
time for the classical evolution of the system $t_i$, with the
corresponding change in the distribution of random initial conditions
(changing by orders of magnitude), as determined by the quantum linear
evolution of the Higgs modes.  Our results are robust to these
changes, as well as to the specific details in which the initial
conditions of the gauge fields are introduced.

We have studied the classical evolution of the system using lattice
methods for a variety of model parameters. Their choice has been done
judiciously to minimize the effect of the ultraviolet and infrared
cutoffs.  The resulting time evolution of global quantities$-$mean
values of the Higgs and inflaton fields, energy fractions, etc.$-$is
fairly insensitive to the randomly chosen (with a fixed distribution)
initial condition. It is also insensitive to the value of the
ultraviolet and infrared cutoffs, if chosen in the appropriate range.
This is confirmed by the shape of the spectral distribution of the
gauge and Higgs fields. The dynamics is dominated by the modes lying
far from the edges, where cutoff effects are dominant.  Therefore, the
results in this region of time seem free of the concerns expressed by
Moore~\cite{moore}.

The main part of our work has focused on the study of the evolution
of Chern-Simons number with time. This turns out to be a much more
fluctuating quantity than the aforementioned average values. This has
forced us to accumulate a considerable amount of statistics in order to
draw conclusive results. Finally, our results provide evidence that the
dynamics of the system leads to the generation of Chern-Simons number, a
prerequisite for it to lead to baryon number generation. The mechanism
for this generation is found to be local and stochastic.

In conclusion, our work provides a step toward the study of a new
mechanism for baryogenesis based on the out of equilibrium situation
occurring during the preheating stage in hybrid inflation
scenario~\cite{GGKS}. There is still a long way to go before this idea
can turn into a full-scale proposal for baryogenesis at the electroweak
scale. Perhaps the most important new ingredient required is the
incorporation of a CP-violating term which would translate the
generation of Chern-Simons number into an actual baryon number at the
required rate. We have assumed in Sect.~\ref{Baryogenesis} a concrete
CP-violating operator which is proportional to the baryonic current and
thus induces a chemical potential for the baryon number~\cite{GGKS},
biasing sphaleron transitions toward baryons rather than
antibaryons. In that case, it seems possible to reproduce the observed
baryon asymmetry of the Universe with a reasonable amplitude of the
CP-violating interaction, in the context of sphaleron transitions at EW
symmetry breaking, i.e. during tachyonic preheating after hybrid
inflation.

\section*{Acknowledgements}

It is a pleasure to thank E. Copeland, D. Grigoriev, F. Klinkhamer, A.
Kusenko, G. Moore, A. Rajantie, P. Saffin, M. Shaposhnikov, J. Smit,
and A. Tranberg for useful comments, suggestions and constructive
criticism.  This work was supported in part by a CICYT project
FPA2000-980.

\appendix

%%%%%%%%%%%%%%%%%%%%%%%%%%%%%%%%%%%%
%%%%%%    SECTION APPENDIX A
%%%%%%%%%%%%%%%%%%%%%%%%%%%%%%%%%%%%

\section{The Lattice Equations of Motion}

In this appendix we explain in full detail the derivation of the
lattice equations of motion. We start by fixing our conventions.  Let
us denote the lattice by $\Lam$. The lattice points are labeled
$n=(n_0,\vec{n})$, whose corresponding space-time position is
$x=(x_0=n_0a_t,\vec{x}=a\vec{n})$, where $a_t$ and $a$ are the
temporal and spatial lattice spacings.  The ratio of spacings will be
called $\kappa=a_t/a$. The lattice inflaton and Higgs fields are
$\INFL(n)$ and $\HIGL(n)$. They are dimensionless and correspond to
the product of the continuum fields by the lattice spacing $a$. Gauge
fields are given by link variables $U_{\mu}(n)$, which are SU(2)
matrices. Therefore, both the gauge and Higgs fields are described by
$2\times 2$ matrices of a certain type. If we choose the matrices
$\overline{\sigma}_{\alpha}\equiv(\uno, i \vec{\tau})$ as a basis of
the space of complex $2 \times 2$ matrices ($\vec{\tau}$ are the Pauli
matrices), we might decompose the lattice Higgs field as follows:

\be
\HIGL(n)= \sum_{\alpha} \higl^{\alpha}(n)\, \overline{\sigma}_{\alpha} \,, 
\ee
where the coefficients $\higl^{\alpha}(n)$ are real. Matrices of this
type are closed under addition and multiplication. They form a field
$\FQ$, which is isomorphic to the field of quaternions. The link
variables $U_{\mu}(n)$ are also members of this space: 
\be\label{cdef}
U_{\mu}(n)=\sum_{\alpha} c_{\mu\alpha}(n) \, \overline{\sigma}_{\alpha} \,, 
\ee
where the coefficients are real and define, for every direction and
lattice point, a vector of unit modulus:
\be\label{cconst}
\sum_{\alpha}c^2_{\mu\alpha}(n)=1 \,.  
\ee 
An arbitrary matrix
${\mathbf B}$ belonging to $\FQ$ satisfies the following relations:
\bea
{\mathbf B}^{\dagger} &\EMI& -{\mathbf B}\\
{\mathbf B}^{\dagger} {\mathbf B}\ &\EMI& 0 \,, 
\eea
where  $\EMI$ stands for equality up to a multiple of the
$2\times 2$ unit matrix.  We will make use of the previous relations
extensively in the following.

A lattice gauge transformation is given by a collection of SU(2)
matrices, one for every point of space $\Omega(n)$. The fields
transform as follows:
\bea
\HIGL(n) &\longrightarrow& \Omega(n)\, \HIGL(n)\,, \\
U_{\mu}(n) &\longrightarrow& \Omega(n)\, U_{\mu}(n)\, 
\Omega^{\dagger}(n+\hat{\mu})
\,, 
\eea 
where $\hat{\mu}$ is the unit vector in the $\mu$ direction.  We will
also be concerned about composite lattice fields defined on
plaquettes:
\be 
P_{\mu \nu}(n)=U_{\mu}(n)\,
U_{\nu}(n+\hat{\mu})\,
U^{\dagger}_{\mu}(n+\hat{\nu})\,U^{\dagger}_{\nu}(n)\,, 
\ee 
which transform as 
\be 
P_{\mu \nu}(n) \ \longrightarrow\ \Omega(n) P_{\mu
  \nu}(n) \Omega^{\dagger}(n)\,.  
\ee

It will be convenient to introduce lattice covariant derivatives
$\DCOV_{\mu}$. The lattice covariant derivative is different for
fields transforming differently under gauge transformations. However,
for ease of notation we will use the same symbol for all types of
fields. In any case, the covariant derivative of a field transforms
like the field itself.  For a Higgs field the covariant derivative is
given by
\be
(\DCOV_{\mu}\, \HIGL)(n) = U_{\mu}(n)\, \HIGL(n+\hat{\mu}) - \HIGL(n) \,.
\ee
For a link variable and a plaquette variable the covariant derivatives are
given by
\bea
&(\DCOV_{\mu}\, U_{\rho})(n) = U_{\mu}(n)\, U_{\rho}(n+\hat{\mu})\,
U_{\mu}^{\dagger}(n+\hat{\rho}) - U_{\rho}(n)\,,  \\
&(\DCOV_{\mu}\, P_{\rho \nu})(n) = U_{\mu}(n)\, P_{\rho
\nu}(n+\hat{\mu})\,U_{\mu}^{\dagger}(n)-  P_{\rho \nu}(n) \,. 
\eea
For vanishing  gauge fields  the 
lattice covariant
derivative reduces to the forward difference operator $\Delta_{\mu}$:
\be
(\Delta_{\mu} f)(n) = f(n+\hat{\mu})-f(n)\,.
\ee
We will also need an adjoint covariant derivative operator $\DBCOV_{\mu}$
reducing to the backward difference operator $\overline{\Delta}_{\mu}$:
\be
(\overline{\Delta}_{\mu} f)(n)=f(n-\hat{\mu})-f(n)\,.
\ee
For a Higgs field one has
\be
(\DBCOV_{\mu}\, \HIGL)(n)= U_{\mu}^{\dagger}(n-\hat{\mu})\,
\HIGL(n-\hat{\mu}) -\HIGL(n)\,.
\ee
The reader can easily work out the corresponding definitions for 
link and plaquette fields.

A final notational convention will help us simplify the lattice
formulas.  We will introduce a lattice metric tensor $\eta_L^{\mu\nu}$
which can be used to raise four-dimensional indices in the standard
way. The tensor is diagonal and its components are $\eta_L^{0 0}=
1/\kappa$, $\eta_L^{i i}= -\kappa$.  This will allow us to use a
restricted Einstein summation convention: Except when explicitly
specified, summation is implied whenever a term contains an upper and
a lower space-time index which are identical.

We are now ready to introduce the dynamics of the fields. This is given 
by the lattice action $S_L$:
\bea \nonumber
S_L &=& \sum_{n \in \Lam}\ \left[ L_G(n)+L_H^{(0)}(n)+ 
L_I^{(0)}(n) \right.\\
&& -\,\kappa\, V[\HIGL(n),\INFL(n)]\Big] \,. 
\eea
The Lagrangian has been split into four parts. $L_G$ contains the pure
gauge interaction, $L_H^{(0)}$ and $L_I^{(0)}$ are the (kinetic) part
of the total Lagrangian containing space-time (covariant) derivatives
of the Higgs and inflaton fields, respectively; finally, the potential
$V(n)\equiv V[\HIGL(n),\INFL(n)]$ contains only point-like interactions
of these fields. The explicit forms of the different Lagrangian terms
that we are using are
\bea \nonumber
L_G(n)&=&{2\over\kappa\gw^2}\,\sum_i \trace[\uno-P_{0 i}(n)] \\ 
&& -\,{\kappa\over\gw^2}\,\sum_{i\ne j} \trace[\uno-P_{i j}(n)]\,,\\
L_H^{(0)}(n)&=& \trace\left[(\DCOV_{\mu}\HIGL)^{\dagger}(n)\; 
(\DCOV^{\mu}\HIGL)(n)\right] \,,\\
L_I^{(0)}(n)&=& \half (\Delta_{\mu}\INFL)(n)\;
(\Delta^{\mu}\INFL)(n) \,,\\ \nonumber
V(n)&=& -\,M_L^2\,\trace[\HIGL^{\dagger}(n)\HIGL(n)] \\
&+& \lambda\, \nonumber
\left(\trace[\HIGL^{\dagger}(n)\HIGL(n)]\right)^2 
+ {\mu_L^2\over2}\INFL^2(n) \\
&+& g^2\,\INFL^2(n)\,
\trace[\HIGL^{\dagger}(n)\HIGL(n)] \,,
\eea
where we have used the lattice metric tensor to raise indices. All
terms are obviously gauge invariant. The constants $g$ and $\lambda$
appearing in the potential are the same couplings that appear in the
continuum Lagrangian.  The dimensionless lattice mass terms $M_L$ and
$\mu_L$ are the product of the continuum mass terms by the lattice
spacing ($M_L=aM$, $\mu_L=a \mu$).

To derive the equations of motion one has to impose that the
derivative of the action with respect to each of the fields vanishes.
Let us begin with the inflaton field $\INFL(n)$. Only the third and
fourth terms in the Lagrangian depend on this field. After
straightforward operations, one obtains
\be\label{INFEQN}
(\Delta_{\mu}\, \overline{\Delta}^{\mu}\, \INFL)(n)=\kappa\left\{\mu_L^2 
+ 2 g^2 \trace[\HIGL^{\dagger}(n)\,\HIGL(n)]\right\} \INFL(n)\,.
\ee
Similarly the equation for the Higgs field can be deduced. In matrix
notation the equation reads:
\be\label{HIGEQN}
(\DCOV_{\mu}\, \DBCOV^{\mu}\, \HIGL)(n)=\kappa\,\Big\{- M_L^2 + 
g^2\INFL^2(n) 
+ 2 \lambda\,\trace[\HIGL^{\dagger}(n)\,
 \HIGL(n)]\Big\}\,\HIGL(n)  \,. 
\ee
In our notation, both equations are very similar to their continuum 
counterparts [see Eq.~(\ref{class_eqs}].

The derivation of the equation of motion for the gauge field is a bit
more tricky. We can deduce it by equating to zero the derivatives of
$L_G+L_H^{(0)}$ with respect to the real coefficients $c_{\alpha\mu}
(n)$, appearing in Eq.~(\ref{cdef}). However, the coefficients $c_{\mu
  \alpha}(n)$ are not independent, since they are subject to the
constraint Eq.~(\ref{cconst}). This can be taken into account by using
Lagrange multipliers. Hence, the equation of motion for the gauge
field has the form
\be
\label{firstge}
\frac{\partial L_G}{\partial c_{\alpha\mu}(n)} + {\partial L_H^{(0)}\over
\partial c_{\alpha\mu}(n)}  = \xi_{\mu}(n)\,  c_{\alpha\mu}(n) \,,
\ee 
where the coefficients $\xi_{\mu}(n)$ are Lagrange multiplier fields,
which are determined by the condition that the evolution preserves the
constraint Eq.~(\ref{cconst}). To work out the first term in
Eq.~(\ref{firstge}), it is convenient to express $L_G$ as follows:
\be
L_G= \trace\left[W^{\mu \dagger}(n)\,U_{\mu}(n)\right] + \ldots \,,
\ee
where $W^{\mu \dagger}(n)$ is given by the standard sum over {\em
  staples}, and the ellipsis represents terms that do not depend on
$U_{\mu}(n)$.  Then, one can reexpress the gauge field evolution
equation as (no summation on $\mu$ on the right hand side)
\be
\label{secondge}
W^{\mu}(n)+K^{\mu}(n) = \xi'_{\mu}(n)\,U^{\mu}(n)\,,   
\ee
where all three terms transform in the same way (as a link variable) under
gauge transformations. The $\xi'$ are not necessarily the same as the
$\xi$ multipliers, but this will be irrelevant in the following. The 
explicit expressions for $K^{\mu}(n)$ and $W^{\mu}(n)$ are
\bea
K^{\mu}(n) &=& (\DCOV^{\mu} \HIGL)(n)\, \HIGL^{\dagger}(n+\hat{\mu})\,,\\
W^{\mu}(n) &=& - \frac{1}{\kappa \gw^2}\, \sum_{\alpha \ne \mu}
\DCOV^{\alpha} \DBCOV_{\alpha}\, U^{\mu}(n) \,.
\eea
Despite the simple expression of our gauge field evolution
equation~(\ref{secondge}), it is convenient to transform it into a form
which more closely resembles the continuum equations of motion.
For that purpose, we multiply the equation by $U_{\mu}^{\dagger}(n)$
and obtain
\be
\label{eqnint}
W^{\mu}(n)\, U_{\mu}^{\dagger}(n) +K^{\mu}(n)\, U_{\mu}^{\dagger}(n) 
\EMI 0 \,,
\ee
where no summation over $\mu$ is implied, and the equality holds
modulo a multiple of the identity matrix. To turn it into an exact
identity we might simply add up the traceless part of the two terms on
the left hand side. For a matrix ${\mathbf B}$ belonging to $\FQ$, the
traceless part is simply $({\mathbf B}-{\mathbf B}^{\dagger})/2$.
Hence, let us introduce the following traceless matrices:
\bea
\label{currentL}
J_L^{\mu}(n)&=& {i\gw\over2}\,\left[K^{\mu}(n)\, U_{\mu}^{\dagger}(n)
-U_{\mu}(n)\, K^{\mu \dagger}(n)\right] \,,\\ \label{Sigma}
\Sigma_L^{\mu}(n)&=& {i\gw\over2}\,\left[W^{\mu}(n)\, U_{\mu}^{\dagger}(n)
-U_{\mu}(n)\, W^{\mu \dagger}(n)\right] \,,
\eea
where again no summation is implied.  The equation of motion for the
gauge fields then reads:
\be
\label{thirdge}
\Sigma_L^{\mu}(n) + J_L^{\mu}(n) = 0 \,.
\ee
The second term in this equation  is the lattice version of the current.
Manipulating Eq.~(\ref{currentL}) one can rewrite it in a way completely 
analogous to the continuum case:
\be
J_L^{\mu}(n)=\frac{i \gw}{2}\,\left[\HIGL(n)\, 
(\DCOV^{\mu} \HIGL)^{\dagger}(n) -
(\DCOV^{\mu} \HIGL)(n)\, \HIGL^{\dagger}(n)\right]\,. 
\ee
To cast Eq.~(\ref{Sigma}) in a way similar to the continuum expression
[Eq.~(\ref{class_eqs})  of Sect. \ref{Methodology}] we introduce the 
lattice counterpart of the field strength:
\be
{\cal F}_{\alpha \beta}(n)= \frac{i}{ 2 \gw }\,\left[P_{\alpha
\beta}(n)-P_{\beta \alpha}(n)\right]\,.
\ee
The gauge equation of motion can finally be written as
\be\label{GAUEQN}
\frac{1}{\kappa}\, \DBCOV_{\alpha}\, {\cal F}^{\mu
\alpha}(n)=J_L^{\mu}(n)\,. 
\ee

In all the previous derivation we have worked in an arbitrary gauge.
However, to have a uniquely defined initial value problem in order to
solve the equations numerically, it is convenient to fix the temporal
gauge,
\be
U_0(n)=\uno \ \Longleftrightarrow \ 
c_{\alpha 0}(n) = \delta_{\alpha 0}\,.
\ee
The lattice equations of motion can now be used to express the value
of the fields at time $n_0+2$ in terms of the fields at time $n_0+1$
and $n_0$.  By using these relations iteratively, one achieves the
numerical evolution of the system. However, although the temporal
component of the gauge potential has been set to zero, the corresponding
equation of motion is still present.  This is precisely Gauss's law:
\be
\label{GaussLaw}
\frac{1}{\kappa}\, \DBCOV_{k} {\cal F}^{0
k}(n)=J_L^{0}(n)\,.    
\ee
This equation is of a different nature than the other ones, since it
relates the field values at times $n_0$ and $n_0+1$: It is a
constraint on the initial value data. Indeed, in the continuum we know
that the constraint is conserved through the evolution equations, so
that if it is satisfied at one time, it is satisfied at all times. Our
next goal will be to show that this is also the case on the lattice.

To study the time variation of Gauss's law, we apply the operator
$\overline{\Delta}_0$ to both sides of Eq.~(\ref{GaussLaw}). Using 
the equation of motion for the Higgs field one obtains
\be
\label{conteq}
\overline{\Delta}_0 J_L^{0}(n)= -\DBCOV_i\, J_L^{i}(n)\,,
\ee
which is the lattice counterpart of the continuity equation. 

Now  we should study the action of the lattice backward time difference 
operator on the left  hand side of Eq.~(\ref{GaussLaw}). The essential 
property that will be needed in the proof is:
\be
 \DBCOV_{\alpha} \DBCOV_{\beta}   {\cal F}^{\alpha \beta} =  
\DBCOV_{\beta} \DBCOV_{\alpha}   {\cal F}^{\alpha \beta}\,,
\ee
where there is no summation implied on either $\alpha$ or $\beta$.
This expresses the commutativity of the covariant derivatives when
acting on the field strength. In general, however, the covariant
derivatives do not commute.  Given this commutativity, we obtain the
following chain of equalities:
\bea \nonumber
\overline{\Delta}_0\DBCOV_{k} {\cal F}^{0 k} =
\DBCOV_{k} \overline{\Delta}_0 {\cal F}^{0 k} &=& 
-\DBCOV_{k} \DBCOV_{i} {\cal F}^{i k} - 
\kappa\,\DBCOV_{k} J_L^{k}(n) \\
&=& - \kappa\, \DBCOV_{k} J_L^{k}(n)\,.
\eea
In the first equality we used the commutativity, in the second the
gauge field equations of motion, and in the third we used the
commutativity again together with the antisymmetry of the field
tensor. Notice that there is now an implicit summation over $i$ and
$k$.

In summary, we have showed that both sides of Gauss's law vary in the
same way with time. This implies not only that fields satisfying
Gauss's law initially will satisfy it at later times, but also that any
initial deviation will remain constant with time. Altogether, the
conclusion is based on the consistency of the gauge field equations
with the covariant conservation of the current, which relies on
\be
 \DBCOV_{\alpha} \DBCOV_{\beta}   {\cal F}^{\alpha \beta} = 0 \,,
\ee  
where now summation is implied.

%%%%%%%%%%%%%%%%%%%%%%%%%%%%%%%%

%%%%%%%%%%%%%%%%%%%%%%%%%%%%%%%%%%%%
%%%%%%    SECTION APPENDIX B
%%%%%%%%%%%%%%%%%%%%%%%%%%%%%%%%%

\section{The Gauss Constraint}

In this appendix we will explain in detail how we have implemented the
Gauss constraint on the initial conditions for the Higgs and gauge
fields. The procedure that we have used in practice is very similar to
the one adopted in \cite{ST} for Abelian gauge fields.

The Gauss constraint relates the electric component of the gauge field
to the current. Thus, our criterion, in agreement with the
considerations made in Sect.~\ref{initial}, has been to fix the
initial (at $t=t_i$) magnetic field to zero. This allows us to fix the
gauge in such a way that the initial vector potential is zero, but
with a nonzero time derivative, i.e. a nonzero electric field
strength.  The resulting form of the Gauss constraint equation in the
$A_0=0$ gauge is the same as in the Abelian case, $\partial_0
\vec\nabla \vec A (\vec x, t_i)= J_0(\vec x, t_i)$, where $J_0$ is the
charge density generated by the Higgs field.  A necessary condition
for this equation to have a solution is that the total charge
vanishes, i.e.  $Q(t_i)=\int d\vec x J_0 (\vec x,t_i) =0$.  This
condition is automatically satisfied with our particular choice of
initial conditions for the Higgs fields, since $ \partial_0 \phi
(k,t_i) = C(k,t_i) \phi(k,t_i)$ with $C(k,t_i)$ a real
color-independent constant. To obtain a unique solution we can impose
the requirement that the transverse components of the initial electric
field vanish. This amounts to imposing the condition that not only
the magnetic field, but also its time derivative vanishes at $t=t_i$.
With this choice we obtain the solution of the Gauss constraint in
Fourier space, given by $ \partial_0 \vec A(\vec k,t_i) = \vec k
J_0(\vec k,t_i) / |\vec k|^2$ for $k\ne 0$ and $ \partial_0 \vec
A(\vec k=\vec 0,t_i) =0$.

The previous construction can easily be translated to the lattice. The
spatial links at the initial time are set to unity, and the links at
the next temporal slice are obtained as $\exp\{-i a \gw A_i(n)\}$,
where the Fourier components of $A_i(n)$ follow a formula equivalent
to that of the continuum. This provides a solution to the lattice
Gauss constraint up to terms of order $a^3 A_i^3(n)$, which, due to
the smallness of the initial gauge field, are zero to machine
precision.  As mentioned before, due to rounding errors, small
violations of the Gauss constraint at later times are induced during
the numerical integration of the equations of motion. However, within
the time scales we have analyzed they remain smaller than $10^{-11}$.

\begin{figure}[htb]
\vspace*{8.5cm}
\includegraphics{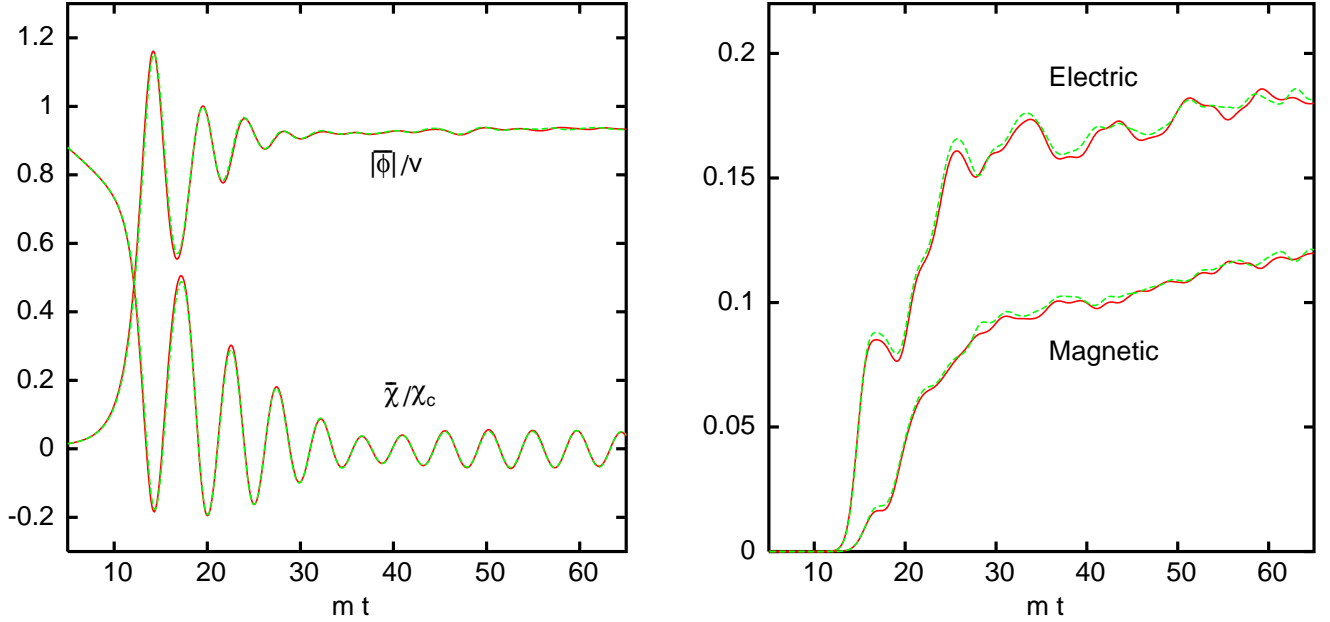}
\caption{ Left: Time evolution of the normalized spatial averages $ \overline{|\phi|}$ and
  $\overline{\chi} $ for a single configuration of model A1. Right:
  Time evolution of the electric and magnetic energy fractions (model
  A1).  Solid or dashed lines describe the evolution for the two
  different ways of implementing the Gauss constraint described in the
  text. Solid lines correspond to setting to zero the initial vector
  potential at $t=t_i$ with a nonzero time derivative, keeping the
  Higgs initial distribution unchanged. For dashed lines both the
  gauge vector potential and its time derivative at $t=t_i$ have been
  set to zero, while the Higgs field has been adjusted to satisfy the
  Gauss constraint.}
\label{gaussc}
\end{figure}

The strategy followed in our implementation of the Gauss constraint
has been to avoid modifications in the initial distribution of the
Higgs field. However, there are alternative approaches that, within a
similar spirit, allow us to check the robustness of our choice. In
particular, we could set to zero both the initial electric and
magnetic fields and allow for minimal modifications of the Higgs field
that preserve $|\phi(\vec x , t_i)|$ and $\partial_0| \phi(\vec x ,
t_i)|$. For the Gauss constraint to be satisfied in that case, the
Higgs field has to be adjusted such as to have local zero charge
density: $J_0(\vec x, t_i)=0$.  A particular solution is to take
$\partial_0 [\Phi(\vec x,t_i)/|\phi(\vec x,t_i)|]=0$.  Figure
\ref{gaussc} compares the spatial averages of $|\phi(t,\vec{x})|$,
$\chi(t,\vec{x})$, and the electric and magnetic average energy
fractions thus obtained with the ones derived with our standard
procedure.  The results correspond to two configurations with the same
initial distribution for the Higgs field up to the modifications
induced by the two different ways of imposing the Gauss constraint. As
can be observed the results are rather insensitive to the particular
choice of initial conditions.  The differences observed are
quantitatively similar to the ones obtained from different
realizations of the random Gaussian initial conditions for the Higgs
field.

\end{document}